\documentclass{aa_v7.0}
\usepackage[varg]{txfonts}
\usepackage{longtable}
\usepackage{graphicx}
\newcommand{\kmprs}  {\mbox{\rm km\,s$^{-1}$}}
\newcommand{\feh} {\mbox{\rm [Fe/H]}}

\newcommand{\xh} {\mbox{\rm [X/H]}}

\newcommand{\oh} {\mbox{\rm [O/H]}}
\newcommand{\co} {\mbox{\rm [C/O]}}
\newcommand{\xfe} {\mbox{\rm [X/Fe]}}
\newcommand{\cfe} {\mbox{\rm [C/Fe]}}
\newcommand{\ofe} {\mbox{\rm [O/Fe]}}
\newcommand{\nafe} {\mbox{\rm [Na/Fe]}}
\newcommand{\mgfe} {\mbox{\rm [Mg/Fe]}}
\newcommand{\alfe} {\mbox{\rm [Al/Fe]}}
\newcommand{\sife} {\mbox{\rm [Si/Fe]}}
\newcommand{\sfe} {\mbox{\rm [S/Fe]}}
\newcommand{\cafe} {\mbox{\rm [Ca/Fe]}}
\newcommand{\tife} {\mbox{\rm [Ti/Fe]}}
\newcommand{\crfe} {\mbox{\rm [Cr/Fe]}}

\newcommand{\nife} {\mbox{\rm [Ni/Fe]}}

\newcommand{\znfe} {\mbox{\rm [Zn/Fe]}}
\newcommand{\yfe} {\mbox{\rm [Y/Fe]}}

\newcommand{\ymg} {\mbox{\rm [Y/Mg]}}
\newcommand{\alphafe} {\mbox{\rm [$\alpha$/Fe]}}
\newcommand{\teff}  {\mbox{$T_{\rm eff}$}}
\newcommand{\Tc}  {\mbox{$T_{\rm C}$}}
\newcommand{\logteff} {\mbox{${\rm log}\,T_{\rm eff}$}}
\newcommand{\logg}  {\mbox{{\rm log}\,$g$}}
\newcommand{\turb}  {\mbox{$\xi_{\rm turb}$}}

\newcommand{\CI} {\ion{C}{i}}
\newcommand{\OI} {\ion{O}{i}}
\newcommand{\oI} {\mbox{\rm [{\ion{O}{i}}]}}

\newcommand{\MgI} {\ion{Mg}{i}}
\newcommand{\NaI} {\ion{Na}{i}}
\newcommand{\AlI} {\ion{Al}{i}}
\newcommand{\SiI} {\ion{Si}{i}}
\newcommand{\SI} {\ion{S}{i}}

\newcommand{\CaI} {\ion{Ca}{i}}
\newcommand{\TiI} {\ion{Ti}{i}}
\newcommand{\TiII} {\ion{Ti}{ii}}
\newcommand{\CrI} {\ion{Cr}{i}}
\newcommand{\CrII} {\ion{Cr}{ii}}

\newcommand{\FeI} {\ion{Fe}{i}}
\newcommand{\FeII} {\ion{Fe}{ii}}
\newcommand{\NiI} {\ion{Ni}{i}}

\newcommand{\ZnI} {\ion{Zn}{i}}

\newcommand{\YII} {\ion{Y}{ii}}

\def\ltsima{$\; \buildrel < \over \sim \;$}
\def\simlt{\lower.5ex\hbox{\ltsima}}
\def\gtsima{$\; \buildrel > \over \sim \;$}
\def\simgt{\lower.5ex\hbox{\gtsima}}
\begin{document}

\title{High-precision abundances of elements in solar twin stars
\thanks{Based
on data products from observations made with ESO Telescopes
at the La Silla Paranal Observatory under programs given in 
Table \ref{table:obs}. 
Tables 2 and 3 are available in electronic form at 
{\tt http://www.aanda.org}.}}

\subtitle{Trends with stellar age and elemental condensation temperature}

\author{P. E. Nissen \inst{1}}

\institute{Stellar Astrophysics Centre, 
Department of Physics and Astronomy, Aarhus University, Ny Munkegade 120, DK--8000
Aarhus C, Denmark.  \email{pen@phys.au.dk}.}

\date{Received 7 April 2015 / Accepted ......}

\abstract
% context heading (optional)
{High-precision determinations of abundances of elements in the atmospheres of the Sun and 
solar twin stars indicate that the Sun 
has an unusual low ratio between refractory and volatile elements. This has led
to the suggestion that the relation between abundance ratios, [X/Fe], and
elemental condensation temperature, \Tc , can be used as a signature of the
existence of terrestrial planets around a star.}
% aims heading (mandatory)
{HARPS spectra with $S/N\simgt600$ for 21 solar twin stars in the solar
neighborhood and the Sun (observed via reflected light from asteroids)
are used to determine 
very precise ($\sigma\!\sim0.01$\,dex) differential abundances of elements 
in order to see how well [X/Fe] is correlated with \Tc\ and
other parameters such as stellar age.} 
% methods heading (mandatory)
{Abundances of C, O, Na, Mg, Al, Si, S, Ca, Ti, Cr, Fe, Ni, Zn,  and Y
are derived from
equivalent widths of weak and medium-strong spectral lines using
MARCS model atmospheres with parameters determined from the excitation and ionization
balance of Fe lines. Non-LTE effects are considered and taken into account for some
of the elements. In addition, precise ($\sigma\!\simlt0.8$\,Gyr)
stellar ages are obtained by interpolating
between Yonsei-Yale isochrones in the \logg -\teff\ diagram.} 
% results heading (mandatory)
{It is confirmed that the ratio between refractory and volatile elements
is lower in the Sun than in most of the solar twins (only
one star has the same \xfe -\Tc\ distribution as the Sun), but for
many stars, the relation between \xfe\ and \Tc\ is not well defined.
For several elements there is an astonishingly tight correlation 
between \xfe\ and stellar age with amplitudes up to $\sim\!0.20$\,dex over an
age interval of eight Gyr in contrast to the lack of correlation between
\feh\ and age. While \mgfe\ increases with age,
the $s$-process element yttrium shows the opposite behavior
meaning that \ymg\ can be used as a sensitive chronometer for Galactic
evolution. \nafe\ and \nife\ are not well correlated with
stellar age, but define a tight Ni-Na relation similar to that 
previously found for more metal-poor stars  albeit with a smaller amplitude.
Furthermore, the C/O ratio evolves very little with time, although \cfe\
and \ofe\ change by $\sim\!0.15$\,dex.}
% conclusions heading (optional)
{The dependence of \xfe\ on stellar age and the \nife - \nafe\ variations
complicate the use of the \xfe -\Tc\ relation as a possible
signature for the existence of terrestrial planets around stars.
The age trends for the various abundance ratios provide
new constraints on supernovae yields and Galactic
chemical evolution, and the slow evolution of C/O for solar metallicity
stars is of interest for discussions of the composition of exoplanets.}


\maketitle

\newpage

\section{Introduction}
\label{sect:introduction} 

In an important paper based on very precise determinations 
of abundances in the Sun and 11 solar twin stars, 
Mel\'{e}ndez et al. (\cite{melendez09}) found that the Sun has a 
lower refractory to volatile element ratio than most
of the solar twins. They suggest that this may be explained by the 
sequestration of refractory elements in terrestrial planets 
and that the slope of abundances relative to iron, 
[X/Fe]\,\footnote{For two elements, X and Y,
with number densities $N_{\rm X}$ and $N_{\rm Y}$,
[X/Y] $\equiv {\rm log}(N_{\rm X}/N_{\rm Y})_{\rm star}\,\, - 
\,\,{\rm log}(N_{\rm X}/N_{\rm Y})_{\rm Sun}$.}, as a function of
elemental condensation temperature, \Tc , in a solar-composition
gas (Lodders \cite{lodders03}) can be used as a signature of the
existence of terrestrial planets around a star. This idea was supported by
Ram\'{\i}rez et al. (\cite{ramirez09}), who in a study of abundances
of 64 solar-type stars find significant variations of the \xfe -\Tc\
slope with only about 15\% of the stars having a slope similar to that of the Sun.
%Hence, stars with terrestrial planets may be relatively rare 
%among solar-type stars, if the Mel\'{e}ndez - Ram\'{\i}rez hypothesis
%is correct.
 
The works of Mel\'{e}ndez et al. (\cite{melendez09}) and 
Ram\'{\i}rez et al. (\cite{ramirez09}) have triggered several
papers investigating relations between  \xfe -\Tc\ slopes and 
existence of planets around stars (Gonzalez et al. \cite{gonzalez10};
Gonz\'{a}lez Hern\'{a}ndez et al. \cite{gonzalez.her10}, \cite{gonzalez.her13};
Schuler et al. \cite{schuler11}; Adibekyan et al. \cite{adibekyan14};
Maldonado et al. \cite{maldonado15}).
The results obtained are not so conclusive, partly because
the stars included span larger ranges in \teff , \logg , and \feh\
than the solar twins in Mel\'{e}ndez et al. (\cite{melendez09})
making the abundance ratios less precise and introducing a dependence
of the derived \xfe -\Tc\ slopes on possible corrections for
Galactic evolution effects. Thus, Gonz\'{a}lez Hern\'{a}ndez et al. (\cite{gonzalez.her13}) 
obtain both positive and negative slopes for ten stars with detected super-Earth
planets, and Adibekyan et al. (\cite{adibekyan14}) find evidence that the slope
depends on stellar age. Furthermore, Schuler et al. (\cite{schuler11}) show
that the  \xfe -\Tc\ slope depends critically on whether one consider
refractory elements only ($\Tc > 900$\,K) or include also volatile elements
with $\Tc < 900$\,K. 

Sequestration in terrestrial planets is not the
only possible explanation of the low abundances of refractory
elements in the Sun. \"{O}nehag et al. (\cite{onehag14})
found that the \xfe -\Tc\ trend for solar-type
stars belonging to the open cluster M\,67 agrees with the trend
for the Sun and suggest that the gas of the proto-cluster was depleted
in refractory elements
by formation and cleansing of dust before the stars formed. According to this, the
Sun may have formed in a dense stellar environment contrary to most
of the solar twins. Similar scenarios are discussed by Gaidos (\cite{gaidos15}),
who suggests that \xfe -\Tc\ correlations can be explained by dust-gas
segregation in circumstellar disks  rather than formation of terrestrial
planets.

Additional information on the connection between abundance ratios in
stars and the occurrence of planets may be obtained from studies of wide
binaries for which the two components have similar \teff\ and \logg\
values. Dust-gas segregation before planet formation
is likely to affect the chemical composition of the components in the
same way. In the case of \object{16 Cyg}\,A/B, for which the secondary component
has a detected planet of at least 1.5 Jupiter mass ($M_{\rm J}$), 
there is a difference \feh (A--B)\,$\simeq\!0.04$\,dex (Ram\'{\i}rez et al.
\cite{ramirez11}) and a difference in the \xfe -\Tc\ slope for refractory
elements (Tucci Maia et al. \cite{tuccimaia14}), which may be explained
by accretion of elements in the rocky core of the giant planet around
\object{16 Cyg\,B}. Furthermore, the two components of \object{XO-2}, which
are hosts of different planets, have an
abundance difference of $\sim\!0.05$\,dex for refractory elements
(Teske et al. \cite{teske15}; Damasso et al. \cite{damasso15}).
On the other hand, Liu et al. (\cite{liu14}) did not
detect any abundance differences between the two components of 
\object{HAT-P-1}, for which the secondary has a $0.5 M_{\rm J}$
planet.

Recent precise determinations of abundances in four solar twin stars:
\object{HD101364} (Mel\'{e}ndez et al. \cite{melendez12}), \object{HIP\,102152} 
(\object{HD197027}) (Monroe et al. \cite{monroe13}), 
\object{18\,Sco} (\object{HD\,146233}) (Mel\'{e}ndez et al. \cite{melendez14a}), and
\object{HD218544} (Mel\'{e}ndez et al. \cite{melendez14b}) 
show that the first three mentioned  stars have a similar depletion 
of refractory elements as the Sun, whereas \object{18\,Sco} 
has a \xfe -\Tc\ slope for elements from carbon to zinc similar 
to the majority of solar twin stars. Neutron capture elements in \object{18\,Sco}
are, however, significantly more 
abundant than lighter elements with the same condensation temperature, and
in \object{HIP\,102152} several elements (N, Na, Co, and Ni), have abundance
ratios with respect to Fe that clearly fall below the 
\xfe -\Tc\ trend for the other elements. These anomalies suggest that
the \xfe -\Tc\ relation is not so well defined as assumed in some of the
works cited above, but that
star-to-star variations in abundance ratios related to e.g. 
incomplete mixing of nucleosynthesis products or chemical evolution play a role. 

In order to make further studies of \xfe -\Tc\ relations, 
I have derived very precise 
abundance ratios ($\sigma \xfe\!\sim\!0.01$\,dex) for a sample of 21 
solar twin stars for which high-resolution HARPS spectra with signal-to-noise (S/N)
ratios ranging from 600 to more than 2000 are available. For the Sun a HARPS
spectrum with $S/N \simeq 1200$  of reflected sunlight from Vesta is used. 
The sample is not well-suited for studying
relations between abundance ratios and occurrence of planets, 
because only three of the stars are known to be hosts of planets. On the other
hand, it is possible to study how well abundance ratios are 
correlated with \Tc\ and to see if \xfe\ depends on other
parameters such as stellar age.

\section{Observational data}
\label{sect:obs}

Based on the precise effective temperatures, surface gravities and metallicities 
derived by Sousa et al. (\cite{sousa08}) from HARPS spectra of FGK  stars,
a sample of solar twins was selected to have parameters that agree
with those of the Sun within $\pm 100$\,K in \teff, $\pm 0.15$\,dex in \logg, and
$\pm 0.10$\,dex in \feh . Listed in Table \ref{table:obs},
21 of these stars have HARPS spectra with $S/N \ge 600$   
after combination of spectra available in the  ESO Science Archive.  

The HARPS spectra have  a resolution of $R\simeq \! 115\,000$
(Mayor et al. \cite{mayor03}). 
Individual spectra for a given star 
were combined after correction for Doppler shifts and then normalized
with the IRAF {\tt continuum} task using cubic splines with a 
scale length of $\sim\!100$\,\AA\ in the blue spectral region (3800 - 5300\,\AA )
and  $\sim\!15$\,\AA\ in the red region (5350 - 6900\,\AA ), for which the 
archive spectra are more irregular.
The S/N values given in Table \ref{table:obs}
were estimated from the rms variation of the flux in
continuum regions for the 4700 - 6900\,\AA\
spectral region, which contains the spectral lines applied in this study.
By including spectra obtained from
2011 to 2013 (programs 183.C-0972 and 188.C-0265), the S/N 
is in most cases significantly higher than that
of the HARPS spectra applied in previous studies of solar twin stars
(e.g., Gonz\'{a}lez Hern\'{a}ndez et al. \cite{gonzalez.her10}, \cite{gonzalez.her13}).
For the stars in Table \ref{table:obs}, the S/N ranges from 600 to 2400 with
a median value of 1000.

The IRAF {\tt splot} task was used to
measure equivalent widths (EWs) of spectral lines by Gaussian fitting 
relative to pseudo-continuum regions lying within 3\,\AA\
from the line measured. These regions
do not necessarily represent the true continuum,
but care was taken to use the same continuum windows in all stars, so
that differences in EW between stars are precisely measured given that the
same instrument and resolution were applied for obtaining spectra. In this
way differential abundances may be obtained with a precision better than 0.01\,dex
as shown by Bedell et al. (\cite{bedell14}). 

In order to achieve high-precision abundances relative to the Sun, it is
very important that a solar flux spectrum is obtained with the same spectrograph
as applied for the solar twin stars (Bedell et al. \cite{bedell14}). 
As listed in Table \ref{table:obs}, several such HARPS spectra are available
based on reflected light from asteroids and the Jupiter moon  Ganymede. 
As the prime reference spectrum for the Sun, the Vesta spectrum with $S/N \simeq 1200$
is adopted, but it is noted that the Ceres and Ganymede spectra agree very well with that
of Vesta. Equivalent widths measured in the Vesta spectrum and the combined spectrum
of Ceres and Ganymede ($S/N \simeq 700$) have an rms difference of 0.35\,m\AA\ only.

\begin{table}
\caption[ ]{ESO/HARPS observing programs and S/N of combined spectra.}
\label{table:obs}
\setlength{\tabcolsep}{0.30cm}
\begin{tabular}{llr}
\noalign{\smallskip}
\hline\hline
\noalign{\smallskip}
  Star & ESO/HARPS program numbers & S/N \\ 
\noalign{\smallskip}
\hline
\noalign{\smallskip}
    HD\,2071 & 072.C-0488,\, 183.C-0972,\, 188.C-0265             &  1050  \\
    HD\,8406 & 072.C-0488,\, 183.C-0972                           &   700  \\
   HD\,20782 & 072.C-0488,\, 183.C-0972                           &  1250  \\
   HD\,27063 & 072.C-0488                                         &   850  \\
   HD\,28471 & 072.C-0488,\, 183.C-0972                           &   700  \\
   HD\,38277 & 072.C-0488,\, 183.C-0972                           &   900  \\
   HD\,45184 & 072.C-0488,\, 183.C-0972                           &  2400  \\
   HD\,45289 & 072.C-0488,\, 183.C-0972,\, 188.C-0265             &  1250  \\
   HD\,71334 & 072.C-0488,\, 183.C-0972,\, 188.C-0265             &   900  \\
   HD\,78429 & 072.C-0488                                         &  1200  \\
   HD\,88084 & 072.C-0488,\, 183.C-0972                           &   800  \\
   HD\,92719 & 072.C-0488,\, 183.C-0972                           &  1150  \\
   HD\,96116 & 072.C-0488,\, 183.C-0972,\, 188.C-0265             &   600  \\
   HD\,96423 & 072.C-0488,\, 183.C-0972,\, 188.C-0265             &  1200  \\
  HD\,134664 & 072.C-0488                                         &   850  \\
  HD\,146233 & 072.C-0488                                         &   850  \\
  HD\,183658 & 072.C-0488,\, 183.C-0972,\, 188.C-0265             &  1000  \\
  HD\,208704 & 072.C-0488,\, 183.C-0972,\, 188.C-0265             &  1100  \\
  HD\,210918 & 072.C-0488                                         &  1300  \\
  HD\,220507 & 072.C-0488,\, 183.C-0972,\, 188.C-0265             &  1100  \\
  HD\,222582 & 072.C-0488,\, 183.C-0972                           &  1000  \\
       Vesta & 088.C-0323                                         &  1200  \\
       Ceres & 082.C-0357,\, 060.A-9036,\, 060.A-9700             &   500  \\
    Ganymede & 060.A-9036,\, 060.A-9700                           &   450  \\
\noalign{\smallskip}
\hline
\end{tabular}

\end{table}

\begin{figure}
\resizebox{\hsize}{!}{\includegraphics{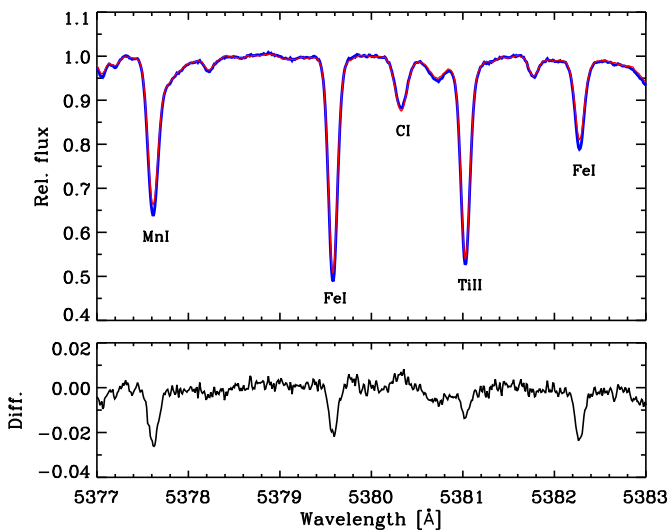}}
\caption{The solar flux HARPS spectrum around the \CI\ line at 5380\,\AA\
obtained via reflected light from Vesta (thin, red line)
in comparison with the HARPS spectrum of \object{HD\,27063} (thick, blue line).
The lower panel shows the difference (HD\,27063 -- Sun) between the two spectra.}
\label{fig:5380}
\end{figure}

To illustrate the high quality of the spectra applied and the differential
way of measuring EWs, Fig. \ref{fig:5380} shows a comparison of the Vesta
spectrum with that of \object{HD\,27063}, which turns out to have a very
similar effective temperature (\teff = 5779\,K) and surface gravity
(\logg = 4.47) as the Sun. As seen from the difference between the spectra
in the lower panel of  Fig. \ref{fig:5380}, lines corresponding to
refractory elements are somewhat stronger in the spectrum of \object{HD\,27063} than in
the solar spectrum, whereas the line of the volatile element carbon at 5380.3\,\AA\
is a bit stronger in the solar spectrum.
\object{HD\,27063} turns out to be a star with a high refractory to volatile ratio
compared to the Sun; i.e. \feh = 0.064 and [C/H] = $-0.018$.

Oxygen is an important volatile element in addition to carbon. Unfortunately,
the HARPS spectra do not cover the \OI\ triplet at 7774\,\AA . Instead, the
forbidden oxygen line at 6300.3\AA\ was used to derive \oh . For six stars
telluric O$_2$ lines happen to disturb the \oI\ line in some of the spectra 
available. Such spectra were excluded before making a combination 
of spectra to be used in the \oI\ region. This decreased the S/N of the 
\oI\ spectra somewhat, but in all cases it is around or above 500.

\section{Model atmosphere analysis}
\label{sect:analysis}

The Uppsala EQWIDTH program, which is based on the assumption of
local thermodynamic equilibrium (LTE), was used to calculate
equivalent widths as a function of element abundance
for plane parallel (1D) MARCS model atmospheres 
(Gustafsson et al. \cite{gustafsson08}) representing
the stars. Interpolation to the observed EWs 
then yields the LTE abundances for the various elements.

\subsection{Spectral lines}
\label{sect:linelist}
The spectral lines used are listed in 
Table \ref{table:linelist}. As seen, they fall in the 
spectral region 4700 - 6900\,\AA , where the S/N of HARPS spectra
is higher and line blending less severe than in the blue region. 
Strong lines, such as the NaD lines and the Mgb triplet, 
and lines significantly blended by stellar or telluric lines
were avoided except for the
\oI\ $\lambda 6300.3$ line, which is blended by a \NiI\ line.
For C, Na, Mg, and Al two lines are available for each element.
Heavier elements have at least three lines and for Ti, Cr, and Fe
both neutral and 
ionized lines are available. In the case of Fe, only lines having
EW$_{\rm Sun} < 70$\,m\AA\ were included. This results in a list  
of 47 \FeI\ lines with excitation potentials, $\chi_{\rm exc}$,
ranging from 0.9 to 5.1 eV 
and nine \FeII\ lines, which are used to determine effective temperatures and surface
gravities of the stars as described in Sect. \ref{sect:parameters}. 

As the analysis is made differentially to the Sun line by line,
the $gf$-values of the spectral lines cancel out. Doppler broadening due to microturbulence
is specified by a depth-independent parameter, \turb .
Collisional broadening of \CaI , \TiI , \CrI , \FeI , \FeII , and \NiI\ lines
caused by neutral hydrogen and helium atoms is based on quantum mechanical 
calculations (Anstee \& O'Mara \cite{anstee95}; Barklem \&  O'Mara \cite{barklem98};
Barklem et al. \cite{barklem00}; Barklem \& Aspelund-Johansson \cite{barklem05}).
For the remaining lines, the Uns{\"o}ld (\cite{unsold55}) approximation with an enhancement
factor of two was applied. If instead an enhancement factor of one is adopted, the 
differential abundances are not changed significantly due to the fact that  
the damping wings of the lines only contribute a small fraction of the equivalent
width.

\onltab{2}{
\clearpage \onecolumn
\begin{longtable}{rccr}
\caption{\label{table:linelist}List of spectral lines.} \\
\hline\hline
\noalign{\smallskip}
  Element & Wavelength & Exc. pot.  & EW$_{\odot}$  \\
          &  [\AA ]    &  [eV]      &     [m\AA ]   \\       
\noalign{\smallskip}
\hline
\noalign{\smallskip}
\endfirsthead
\caption{continued} \\
\hline\hline
\noalign{\smallskip}
  Element & Wavelength & Exc. pot. & $EW_{\odot}$  \\
          &  (\AA )    &  (eV)     &     (m\AA )   \\
\noalign{\smallskip}
\hline
\noalign{\smallskip}
\noalign{\smallskip}
\endhead
\hline
\endfoot
   \CI &   5052.15 &    7.685 &    36.0  \\
   \CI &   5380.32 &    7.685 &    21.4  \\
  \oI  &   6300.31 &    0.000 &     3.6  \\
  \NaI &   6154.23 &    2.102 &    38.2  \\
  \NaI &   6160.75 &    2.104 &    58.8  \\
  \MgI &   4730.04 &    4.340 &    71.5  \\
  \MgI &   5711.10 &    4.345 &   105.6  \\
  \AlI &   6696.03 &    3.143 &    37.7  \\
  \AlI &   6698.67 &    3.143 &    21.0  \\
 \SiI &    5517.55 &    5.080 &    13.8  \\
 \SiI &    5645.62 &    4.929 &    36.6  \\
 \SiI &    5665.56 &    4.920 &    41.2  \\
 \SiI &    5793.08 &    4.929 &    43.9  \\
 \SiI &    6125.03 &    5.614 &    32.5  \\
 \SiI &    6145.02 &    5.616 &    38.8  \\
 \SiI &    6243.82 &    5.616 &    48.1  \\
 \SiI &    6244.48 &    5.616 &    46.3  \\
 \SiI &    6721.84 &    5.862 &    44.7  \\
 \SiI &    6741.63 &    5.984 &    15.9  \\
 \SI &     6045.98 &    7.868 &    18.3  \\
 \SI &     6052.67 &    7.870 &    11.8  \\
 \SI &     6743.58 &    7.866 &     8.6  \\
 \SI &     6757.14 &    7.870 &    18.7  \\
 \CaI &    5260.39 &    2.521 &    32.8  \\
 \CaI &    5512.99 &    2.933 &    88.3  \\
 \CaI &    5581.98 &    2.523 &    96.5  \\
 \CaI &    5590.13 &    2.521 &    92.6  \\
 \CaI &    5867.57 &    2.933 &    25.0  \\
 \CaI &    6166.44 &    2.521 &    71.1  \\
 \CaI &    6455.60 &    2.523 &    57.4  \\
 \TiI &    4913.62 &    1.873 &    51.5  \\
 \TiI &    5113.45 &    1.443 &    27.9  \\
 \TiI &    5219.71 &    0.021 &    28.1  \\
 \TiI &    5295.78 &    1.067 &    12.7  \\
 \TiI &    5490.16 &    1.460 &    22.2  \\
 \TiI &    5739.48 &    2.249 &     8.0  \\
 \TiI &    5866.46 &    1.066 &    48.6  \\
 \TiI &    6091.18 &    2.267 &    15.2  \\
 \TiI &    6126.22 &    1.066 &    22.6  \\
 \TiI &    6258.11 &    1.443 &    52.2  \\
 \TiI &    6261.11 &    1.429 &    48.5  \\
 \TiII &   5211.54 &    2.590 &    33.7  \\
 \TiII &   5381.03 &    1.565 &    60.9  \\
 \TiII &   5418.77 &    1.582 &    49.0  \\
 \CrI &    5214.13 &    3.369 &    17.0  \\
 \CrI &    5238.97 &    2.709 &    16.9  \\
 \CrI &    5247.57 &    0.960 &    83.4  \\
 \CrI &    5272.00 &    3.449 &    23.8  \\
 \CrI &    5287.18 &    3.438 &    11.3  \\
 \CrI &    5296.70 &    0.983 &    93.9  \\
 \CrI &    5348.33 &    1.004 &   100.2  \\
 \CrI &    5783.07 &    3.323 &    31.4  \\
 \CrI &    5783.87 &    3.322 &    44.6  \\
 \CrI &    6661.08 &    4.193 &    13.0  \\
 \CrII &   5237.33 &    4.073 &    53.5  \\
 \CrII &   5246.78 &    3.714 &    16.2  \\
 \FeI &    5295.32 &    4.415 &    29.9  \\
 \FeI &    5373.71 &    4.473 &    63.9  \\
 \FeI &    5379.58 &    3.694 &    61.9  \\
 \FeI &    5386.34 &    4.154 &    33.1  \\
 \FeI &    5466.99 &    3.573 &    35.9  \\
 \FeI &    5522.45 &    4.209 &    44.2  \\
 \FeI &    5546.51 &    4.371 &    52.9  \\
 \FeI &    5560.22 &    4.434 &    52.6  \\
 \FeI &    5577.03 &    5.033 &    11.6  \\
 \FeI &    5618.64 &    4.209 &    50.4  \\
 \FeI &    5636.71 &    3.640 &    19.6  \\
 \FeI &    5650.00 &    5.100 &    36.5  \\
 \FeI &    5651.48 &    4.473 &    19.0  \\
 \FeI &    5661.35 &    4.284 &    23.0  \\
 \FeI &    5679.03 &    4.652 &    60.4  \\
 \FeI &    5705.47 &    4.301 &    38.4  \\
 \FeI &    5855.09 &    4.608 &    22.7  \\
 \FeI &    6079.02 &    4.652 &    46.5  \\
 \FeI &    6082.72 &    2.223 &    35.4  \\
 \FeI &    6093.65 &    4.607 &    31.1  \\
 \FeI &    6096.67 &    3.984 &    37.8  \\
 \FeI &    6151.62 &    2.176 &    50.1  \\
 \FeI &    6157.73 &    4.076 &    62.4  \\
 \FeI &    6165.36 &    4.143 &    45.0  \\
 \FeI &    6173.34 &    2.223 &    68.7  \\
 \FeI &    6188.00 &    3.943 &    48.4  \\
 \FeI &    6226.74 &    3.883 &    29.6  \\
 \FeI &    6240.65 &    2.223 &    49.0  \\
 \FeI &    6270.23 &    2.858 &    52.2  \\
 \FeI &    6271.28 &    3.332 &    24.4  \\
 \FeI &    6380.75 &    4.186 &    52.7  \\
 \FeI &    6392.54 &    2.279 &    17.6  \\
 \FeI &    6498.94 &    0.958 &    47.1  \\
 \FeI &    6597.57 &    4.795 &    44.0  \\
 \FeI &    6625.04 &    1.011 &    15.6  \\
 \FeI &    6703.58 &    2.759 &    37.3  \\
 \FeI &    6705.10 &    4.607 &    47.6  \\
 \FeI &    6710.32 &    1.485 &    16.0  \\
 \FeI &    6713.75 &    4.795 &    21.3  \\
 \FeI &    6725.36 &    4.103 &    17.8  \\
 \FeI &    6726.67 &    4.607 &    47.3  \\
 \FeI &    6733.15 &    4.638 &    26.9  \\
 \FeI &    6739.52 &    1.557 &    12.0  \\
 \FeI &    6793.27 &    4.076 &    13.0  \\
 \FeI &    6806.86 &    2.727 &    34.9  \\
 \FeI &    6810.27 &    4.607 &    50.1  \\
 \FeI &    6843.65 &    4.548 &    61.5  \\
 \FeII &   5414.08 &    3.222 &    27.3  \\
 \FeII &   5425.26 &    3.200 &    41.5  \\
 \FeII &   6084.10 &    3.200 &    21.2  \\
 \FeII &   6149.25 &    3.889 &    36.7  \\
 \FeII &   6247.56 &    3.892 &    53.1  \\
 \FeII &   6369.46 &    2.891 &    18.8  \\
 \FeII &   6416.93 &    3.892 &    40.1  \\
 \FeII &   6432.68 &    2.892 &    41.6  \\
 \FeII &   6456.39 &    3.904 &    63.4  \\
 \NiI &    4953.21 &    3.740 &    56.2  \\
 \NiI &    5010.94 &    3.635 &    49.9  \\
 \NiI &    5643.09 &    4.164 &    15.9  \\
 \NiI &    5805.23 &    4.167 &    41.5  \\
 \NiI &    6086.29 &    4.266 &    44.2  \\
 \NiI &    6108.12 &    1.676 &    65.6  \\
 \NiI &    6130.14 &    4.266 &    22.3  \\
 \NiI &    6176.82 &    4.088 &    64.1  \\
 \NiI &    6177.25 &    1.826 &    14.8  \\
 \NiI &    6204.61 &    4.088 &    22.1  \\
 \NiI &    6378.26 &    4.154 &    32.8  \\
 \NiI &    6643.64 &    1.676 &    94.7  \\
 \NiI &    6767.78 &    1.826 &    79.7  \\
 \NiI &    6772.32 &    3.657 &    50.1  \\
 \ZnI &    4722.16 &    4.030 &    71.6  \\
 \ZnI &    4810.54 &    4.080 &    75.0  \\
 \ZnI &    6362.35 &    5.790 &    20.4  \\
 \YII &    4883.69 &    1.084 &    59.5  \\
 \YII &    5087.43 &    1.084 &    48.7  \\
 \YII &    5200.42 &    0.992 &    38.1  \\
\noalign{\smallskip}
\hline

\end{longtable}
} %end of onltab

The derivation of oxygen abundances impose a particular problem, because 
the $\oI \, \lambda 6300.3$ line is blended by a \NiI\ line with nearly the same
wavelength (Allende-Prieto et al. \cite{prieto01}).
Based on the MARCS solar model, log\,$gf \, = \, -2.11$
(Johansson et al. \cite{johansson03}), 
and a solar nickel abundance of $A \rm{(Ni)}_{\odot} = 6.15$ 
(as determined from 17 weak \NiI\ lines listed in Scott et al. \cite{scott09}),
the equivalent width of the \NiI\ line is calculated to be 1.7\,m\AA .
After subtracting this value from the measured
EW of the \oI\ - \NiI\ blend (5.3\,m\AA ),
one gets EW(\oI )$_{\odot} = 3.6$\,m\AA . For the stars,
the measured EWs of the \oI\ - \NiI\ blend were corrected in the same way
by subtracting the EW of the \NiI\ line as calculated for the model
atmosphere of the star and the [Ni/H] value derived.

\subsection{Stellar parameters}
\label{sect:parameters}

Adopting solar parameters, \teff \,=\,5777\,K, \logg \,=\,4.438, and \turb \,=\,1.0\,\kmprs ,
the corresponding parameters of the stars were determined by requesting that
\feh\ has no systematic dependence on $\chi_{\rm exc}$ and EW of the lines and 
that the mean \feh\ values derived from \FeI\ and \FeII\ lines, respectively, are equal. 
The slope of \feh\ as a function of excitation potential is sensitive to
the effective temperature and the slope of \feh\ versus EW depends on the microturbulence.
The difference of \feh\ derived from \FeI\ and \FeII\ lines depends on the
surface gravity via its effect on the electron pressure in the stellar
atmospheres, but it is also affected by the iron and $\alpha$-element abundances,
because these elements are important electron donors. Therefore, it it is necessary
to make a number of iterations using MARCS models with different parameters
\teff , \logg , \feh , and \alphafe\ to ensure that the excitation and 
ionization balance of Fe lines are fulfilled. These models are obtained from 
two sets of MARCS models: the standard set with
\alphafe = 0 at $\feh \ge 0$ and \alphafe\ rising linearly to 
\alphafe = 0.4 when \feh\ decreases from 0.0 to $-1.0$, and the
$\alpha$-enhanced set with
$\alphafe = 0.4$ at all metallicities. Hence, it it possible to
interpolate to a MARCS model with the \alphafe\ value\footnote{
Defined as $\alphafe = \frac{1}{4} \cdot (\mgfe\ + \sife\ + \cafe\ + \tife )$ in
this paper.} of a star. 
This is of some importance for three $\alpha$-enhanced solar twins in this
paper with $\alphafe \simeq 0.08$. If models with $\alphafe = 0.0$
were adopted for these stars, the derived effective temperature 
and gravity would change by $\Delta \teff \simeq -8$\,K and 
$\Delta \logg \simeq +0.025$\,dex, whereas the effect on \feh\ is only $-0.003$\,dex. 

\begin{figure}
\resizebox{\hsize}{!}{\includegraphics{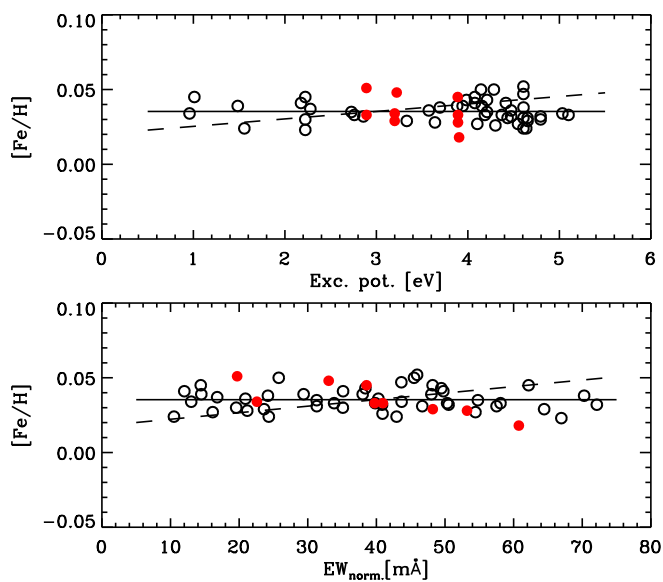}}
\caption{[Fe/H] derived from lines in the spectrum of \object{HD\,183658} as
a function of excitation potential (upper panel) and ``normalized"
equivalent width: $EW_{\rm norm.} = EW \cdot 6000.0 / \lambda$ [\AA ] (lower panel).
Open circles refer to \FeI\ lines and filled (red) circles to \FeII\ lines.
The full drawn lines show linear fits to the \FeI\ data for the derived
parameters of the star. The dashed lines show changes in the slopes
if \teff\ is decreased by 30\,K (upper panel) and the microturbulence
is decreased by 0.1\,\kmprs\ (lower panel).}
\label{fig:183658}
\end{figure}

As a typical example, Fig. \ref{fig:183658} shows \feh\ as a function of
$\chi_{\rm exc}$ and EW obtained with the final model of \object{HD\,183658}
(\teff , \logg, \feh , \alphafe ) = (5809\,K, 4.402, +0.035, 0.007). The
rms scatter of \feh\ is 0.007\,dex for \FeI\ lines and  0.011\,dex for
\FeII\ lines. Based on the 1-sigma errors of the slopes of \feh\ versus 
$\chi_{\rm exc}$ and EW and the errors of the average Fe abundances derived from
\FeI\ and \FeII\ lines, respectively, the error
analysis method described by Epstein et al. (\cite{epstein10}) and
Bensby et al. (\cite{bensby14}) has been used to derive the errors of 
\teff , \logg , \feh, and \turb . The estimated errors are nearly the same for all stars, i.e.,
$\sigma (\teff ) = 6$\,K, $\sigma (\logg ) = 0.012$\,dex,
$\sigma (\feh ) = 0.006$\,dex, and $\sigma (\turb ) = 0.017$\,\kmprs . 

In the derivation of the model atmosphere parameters, small differential
non-LTE corrections for the \FeI\ lines according to Lind et al.
(\cite{lind12}) have been included. The largest non-LTE effects occur at 
the lowest gravities of the solar twin sample, i. e., $\logg \sim 4.25$;
\feh\ is increased by 0.003 to 0.004\,dex and \logg\ by 0.007 to 0.009\,dex
relative to the LTE values.
Changes in \teff\ are negligible, i. e., ranging from +2 to $-3$\,K.

The small errors quoted are only obtainable for samples of solar twin
stars that can be analyzed differentially with respect to the Sun
using the same list of Fe lines. Still, one may wonder if these errors are
really realistic. To test this, I have compared with parameters recently determined for 
88 solar twin stars by Ram\'{\i}rez et al. (\cite{ramirez14b}) based on spectra  
obtained with the 6.5\,m Clay Magellan Telescope and the MIKE spectrograph.
The spectra have $S/N > 400$ and resolutions of $R = 83\,000$ (65\,000) in the
blue (red) spectral regions.  A set of 91 \FeI\ and 19 \FeII\ lines were
analyzed with MARCS models to derive atmospheric parameters with errors similar
to those quoted above. For 14 stars in common with the present paper I get
the following average differences (Ram\'{\i}rez -- this paper) and rms deviations:  
$\Delta \teff = 0 \pm 10$\,K, $\Delta \logg = 0.002 \pm 0.020$,
and $\Delta \feh = 0.000 \pm 0.014$. The rms deviations for \teff\ and \logg\ are
fully explainable by the errors quoted in the two works, whereas the rms deviation
for \feh\ is a bit larger than expected.

The parameters derived in this paper are also in good agreement with those 
derived from HARPS spectra by Sousa et al. (\cite{sousa08})
based on an extensive list of 236 \FeI\ and 36 \FeII\ lines analyzed by using 
a grid of Kurucz model atmospheres (Kurucz \cite{kurucz93}).
The average differences (Sousa -- this paper) and rms deviations 
for the 21 solar twins in the present paper are: 
$\Delta \teff = -1 \pm 8$\,K, $\Delta \logg = 0.018 \pm 0.033$,
and $\Delta \feh = -0.003 \pm 0.009$. The deviations in \teff\ and
\feh\ are compatible with the small errors quoted above, 
but the rms deviation in \logg\ is larger than expected; stars with 
$\logg < 4.35$ turn out to have systematically higher \logg\ values  
in Sousa et al. (\cite{sousa08}) than in the present paper.
These deviations are not seen when comparing with \logg\ values from 
Ram\'{\i}rez et al. (\cite{ramirez14b}). 

\begin{figure}
\resizebox{\hsize}{!}{\includegraphics{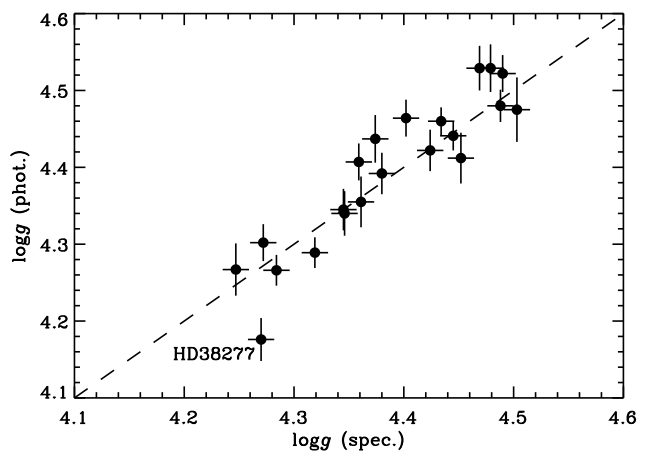}}
\caption{Photometric gravities based on Hipparcos parallaxes
as a function of spectroscopic gravities determined from 
the Fe ionization balance.}
\label{fig:logg.compare}
\end{figure}

The spectroscopic surface gravities have also been compared to photometric
gravities determined from the relation
\begin{eqnarray} 
\log \frac{g}{g_{\odot}}  =  \log \frac{M}{M_{\odot}} +
4 \log \frac{\teff}{T_{\rm eff,\odot}} + 0.4 (M_{\rm bol} - M_{{\rm bol},\odot}),
\end{eqnarray}
where $M$ is the mass of the star and $M_{\rm bol}$ the absolute
bolometric magnitude. Hipparcos parallaxes  
(van Leeuwen \cite{leeuwen07}) were used to derive absolute visual magnitudes
from V magnitudes based on Str\"{om}gren photometry (Olsen \cite{olsen83}), 
and bolometric 
corrections were adopted from Casagrande et al. ({\cite{casagrande10}).
Stellar masses were obtained by interpolating in the luminosity - \logteff\
diagram between the Yonsei -Yale evolutionary tracks of Yi et al. (\cite{yi03});
see Nissen \& Schuster (\cite{nissen12}) for details. The estimated error 
of \logg \,(phot.) ranges from 0.02 to 0.04\,dex depending primarily on the
relative error of the parallax. A seen from Fig. \ref{fig:logg.compare},
there is a satisfactory agreement between the two sets of gravities.
The weighted mean deviation (\logg \,(phot.) -- \logg \,(spec.)) is 0.005\,dex,
and the rms deviation is as expected from the error bars except for
one star, \object{HD\,38277}, which has a 3-sigma deviation. Its lower 
\logg \,(phot.) could be explained if the star is a spectroscopic binary,
but there is no indication of a component in the HARPS spectra.

\subsection{Non-LTE and 3D effects}
\label{sect:non-lte}

Although the solar twin stars have atmospheric parameters close to 
those of the Sun, effects of deviations from LTE may be important
considering the very high precision we are aiming at. 

The detailed calculations by Lind et al. (\cite{lind12}) 
show that \FeII\ lines are not affected by departures from LTE, 
but as mentioned in Sect. \ref{sect:parameters}
there is a small non-LTE effect on Fe abundances 
derived from \FeI\ lines.  Using the IDL program
made available by Lind et al. to calculate non-LTE corrections for the
actual set of \FeI\ lines applied, it is found that
the largest differential correction, $\Delta \feh = 0.004$, occurs for
stars having gravities about 0.15\,dex below
the solar gravity. Calculations of non-LTE corrections as a
function of \teff , \logg , and \feh\ for the lines of \CI\ (Takeda \& Honda 
\cite{takeda05}), \MgI\ (Zhao \& Gehren \cite{zhao00}), \SI\ and \ZnI\
(Takeda et al. \cite{takeda.etal05}), and \CaI\ (Mashonkina et al. \cite{mashonkina07}),
suggest similar  small differential corrections as in the case of the \FeI\
lines;  hence the non-LTE effect on the corresponding abundance
ratios, [X/Fe], is negligible, and the LTE ratio is therefore adopted. 
For the two \NaI\ lines applied, $\lambda 6154.2$ and
$\lambda 6160.7$, the calculations of Lind et al. (\cite{lind11}), indicate
similar numerical corrections as in the case of \FeI\ lines, but with {\em opposite}
sign. Therefore, non-LTE corrections have been applied when determining \nafe\ ratios;
the most extreme correction is $-0.008$\,dex for \object{HD\,38277}. 

For the lines of \AlI\ and \SiI\, there seems to be no 
detailed calculations of non-LTE effects
as a function of \teff , \logg , and \feh\ for solar type stars, but the 
corrections for the Sun itself  are on the order of 0.01\,dex only 
(Baum\"{u}ller \& Gehren
\cite{baumueller96}; Shi et al. \cite{shi08}; Scott et al. \cite{scott15b}).
This suggests that differential corrections for the Al and Si abundances
can be neglected.  It is probably also the case for Ni
(Scott et al. \cite{scott15a}), although extensive non-LTE calculations 
for \NiI\ lines are still missing.

The statistical equilibrium calculations by Bergemann (\cite{bergemann11}) and 
Bergemann \& Cescutti (\cite{bergemann10}) show that the \TiII\  and \CrII\ lines
applied in this paper are formed in LTE in the solar atmosphere, whereas there
are significant non-LTE effects for the \TiI\ and \CrI\ lines.  We may investigate
this empirically by comparing abundances derived from neutral and ionized lines,
respectively. As seen from Fig. \ref{fig:ticr-logg}, there appears to be
a trend of [\ion{Ti}{i}/\ion{Ti}{ii}] as a function of \logg\
suggesting that solar twin stars with the lowest and highest gravities have
differential non-LTE corrections on the order of $\pm 0.01$\,dex
for Ti abundances derived from \TiI\ lines. In view of this, the Ti abundances
derived from \TiII\ lines will be adopted although only three such lines are
available compared to 11 \TiI\ lines. For Cr there is no significant trend
of [\ion{Cr}{i}/\ion{Cr}{ii}] versus \logg\ and the rms scatter 
is only 0.008\,dex. In this case, the Cr abundance derived from \CrI\ lines
is adopted.

\begin{figure}
\resizebox{\hsize}{!}{\includegraphics{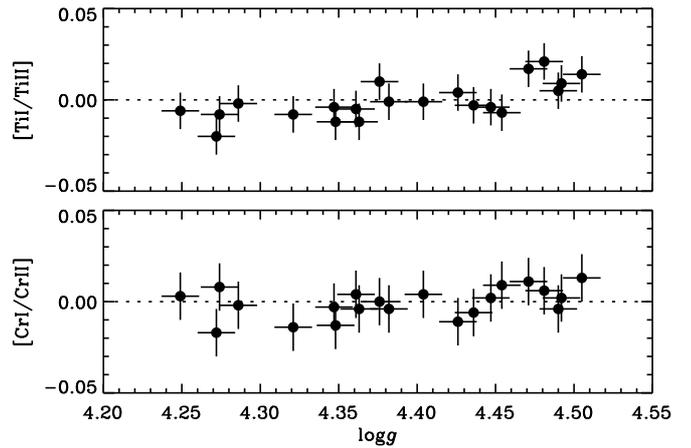}}
\caption{The ratio of Ti abundances derived from \TiI\ and \TiII\ lines,
respectively, as a function of \logg\ (upper panel) and the corresponding 
ratio for  Cr (lower panel).}  
\label{fig:ticr-logg}
\end{figure}

The oxygen abundances derived from the forbidden \oI\ line at
6300\,\AA\ are not affected by departures from LTE (e.g., Kiselman \cite{kiselman93}).
This line is , however, sensitive to 3D hydrodynamical (granulation)
effects on the temperature structure of the stellar atmosphere (Asplund \cite{asplund05}).
According to Nissen et al. (\cite{nissen02}), this leads to 3D corrections
that depend on effective temperature and metallicity. For the present sample
of solar twin stars the largest differential effect on \oh\ is 0.007\,dex.
For the other elements, one would expect  
3D effects to be less significant, because the lines applied are formed deeper
in the atmospheres, but this need to be confirmed by 3D model calculations
preferentially with non-LTE effects included.

The impact of magnetic fields on the temperature structure of stellar atmospheres
is another potential problem. Magneto-hydrodynamic simulations of the
solar atmosphere (Fabbian et al. \cite{fabbian10}, \cite{fabbian12}) show
effects on the strength of \FeI\ lines induced  primarily  by
``magnetic heating" of the upper layers. For an average vertical
magnetic field strength of 100\,G (Trujillo Bueno et al. \cite{trujillo04}),
the effect on the derived 
Fe abundance ranges from 0.01\,dex for weak lines formed in deep layers
to 0.08\,dex for stronger lines formed in the upper part of the photosphere.  
These results have recently been confirmed by Moore et al.
(\cite{moore15}), who in addition find a correction on the order of +0.02\,dex
for the oxygen abundance derived from the \oI\ $\lambda 6300$ line if
a vertical mean field of 80\,G is imposed. A similar correction is
suggested by the recent MHD simulations of Fabbian \& Moreno-Insertis
(\cite{fabbian15}). 

These simulations raise the question if star-to-star differences of the
magnetic field strength
lead to variations of abundance ratios among solar twin stars. If so,
the ratio of Fe abundances determined from \FeI\ and \FeII\ lines
(used for determining the gravity parameter)
would also be affected, because of different depth of formation for
neutral and ionized lines. As discussed in Sect. \ref{sect:parameters},
there is, however, good agreement between spectroscopic gravities and 
gravities determined via Hipparcos parallaxes,
which suggests that differences in magnetic field strength do not play a 
major role in the context of abundance variations among solar twin
stars. Still, this is a problem that deserves further studies.

\subsection{Abundance ratios and uncertainties}
\label{sect:abundances}

The derived abundance ratios, [X/Fe], are given in Table \ref{table:abun}
together with the atmospheric parameters of the stars. 
Due to the systematic deviation
between \logg\ values derived in this paper and those of Sousa et al.
(\cite{sousa08}), four stars have gravities below the selection limit
of \logg \,= 4.29.

\onltab{3}{
\newpage
\begin{table*}
\caption[ ]{Atmospheric parameters and abundance ratios.}
\label{table:abun}
\setlength{\tabcolsep}{0.05cm}
\begin{tabular}{lrrrrrrrrrrrrrrrrr}
\noalign{\smallskip}
\hline\hline
\noalign{\smallskip}
  Star & \teff & \logg & \feh & \turb & \cfe & \ofe & \nafe & \mgfe & \alfe & \sife & \sfe & \cafe & \tife & \crfe & \nife & \znfe & \yfe  \\
  & K &  &  & \kmprs & &  & &  &  &  &  &  &  &  &  &  & \\
\noalign{\smallskip}
\hline
\noalign{\smallskip}
HD2071 &5724 &4.490 &$-$0.084 &  0.96 & $-$0.022 & $-$0.022 & $-$0.032 &  0.001 & $-$0.002 &  0.001 & $-$0.007 &  0.024 &  0.019 &  0.005 & $-$0.029 & $-$0.026 & 0.040 \\
HD8406 &5730 &4.479 &$-$0.105 &  0.95 & $-$0.036 & $-$0.028 & $-$0.059 &  0.000 & $-$0.020 & $-$0.010 & $-$0.021 &  0.031 &  0.009 &  0.010 & $-$0.041 & $-$0.044 & 0.050 \\
HD20782 &5776 &4.345 &$-$0.058 &  1.04 & $-$0.019 &  0.024 & $-$0.086 &  0.011 &  0.015 & $-$0.006 & $-$0.013 &  0.015 &  0.028 & $-$0.004 & $-$0.042 & $-$0.019 & $-$0.110 \\
HD27063 &5779 &4.469 & 0.064 &  0.99 & $-$0.082 & $-$0.078 & $-$0.041 & $-$0.018 & $-$0.029 & $-$0.019 & $-$0.039 &  0.023 & $-$0.005 &  0.016 & $-$0.022 & $-$0.055 & 0.067 \\
HD28471 &5754 &4.380 &$-$0.054 &  1.02 & $-$0.023 &  0.006 & $-$0.013 &  0.037 &  0.065 &  0.022 & $-$0.010 &  0.026 &  0.046 & $-$0.007 & $-$0.005 &  0.018 & $-$0.067 \\
HD38277 &5860 &4.270 &$-$0.070 &  1.17 & $-$0.012 &  0.049 & $-$0.078 &  0.012 &  0.012 &  0.006 & $-$0.017 &  0.007 &  0.023 & $-$0.018 & $-$0.042 & $-$0.016 & $-$0.129 \\
HD45184 &5871 &4.445 & 0.047 &  1.06 & $-$0.069 & $-$0.055 & $-$0.031 & $-$0.009 & $-$0.015 & $-$0.005 & $-$0.045 &  0.015 &  0.011 & $-$0.002 & $-$0.020 & $-$0.030 & 0.040 \\
HD45289\tablefootmark{*} &5718 &4.284 &$-$0.020 &  1.06 &  0.058 &  0.146 & $-$0.009 &  0.100 &  0.128 &  0.055 &  0.056 &  0.049 &  0.100 &  0.002 &  0.006 &  0.096 & $-$0.094 \\
HD71334 &5701 &4.374 &$-$0.075 &  0.98 & $-$0.024 &  0.031 & $-$0.034 &  0.039 &  0.063 &  0.019 & $-$0.003 &  0.034 &  0.041 &  0.000 & $-$0.014 & $-$0.007 & $-$0.076 \\
HD78429 &5756 &4.272 & 0.078 &  1.05 & $-$0.025 &  0.004 & $-$0.055 &  0.022 &  0.034 & $-$0.002 &  0.001 &  0.006 &  0.019 &  0.005 & $-$0.020 &  0.014 & $-$0.101 \\
HD88084 &5768 &4.424 &$-$0.091 &  1.02 &  0.017 & $-$0.010 &  0.003 &  0.028 &  0.053 &  0.021 &  0.008 &  0.019 &  0.035 & $-$0.009 & $-$0.003 &  0.009 & $-$0.044 \\
HD92719 &5813 &4.488 &$-$0.112 &  1.00 & $-$0.060 & $-$0.009 & $-$0.050 &  0.002 & $-$0.014 &  0.003 & $-$0.062 &  0.031 &  0.021 & $-$0.001 & $-$0.044 & $-$0.046 & 0.042 \\
HD96116 &5846 &4.503 & 0.006 &  1.02 & $-$0.088 & $-$0.046 & $-$0.100 & $-$0.041 & $-$0.063 & $-$0.036 & $-$0.052 &  0.019 & $-$0.019 &  0.013 & $-$0.056 & $-$0.085 & 0.095 \\
HD96423 &5714 &4.359 & 0.113 &  0.99 & $-$0.056 & $-$0.076 &  0.008 &  0.021 &  0.056 &  0.005 & $-$0.035 &  0.005 &  0.028 &  0.005 &  0.021 & $-$0.014 & $-$0.033 \\
HD134664 &5853 &4.452 & 0.093 &  1.01 & $-$0.116 & $-$0.087 & $-$0.058 & $-$0.022 & $-$0.035 & $-$0.019 & $-$0.048 &  0.009 &  0.004 &  0.006 & $-$0.022 & $-$0.055 & 0.050 \\
HD146233 &5809 &4.434 & 0.046 &  1.02 & $-$0.066 & $-$0.028 & $-$0.028 & $-$0.010 & $-$0.020 & $-$0.007 & $-$0.029 &  0.010 &  0.009 &  0.010 & $-$0.013 & $-$0.036 & 0.041 \\
HD183658 &5809 &4.402 & 0.035 &  1.04 & $-$0.002 & $-$0.040 &  0.024 &  0.009 &  0.018 &  0.009 &  0.006 &  0.002 &  0.008 & $-$0.005 &  0.013 &  0.008 & $-$0.025 \\
HD208704 &5828 &4.346 &$-$0.091 &  1.08 & $-$0.009 &  0.033 & $-$0.062 &  0.017 &  0.023 &  0.005 & $-$0.020 &  0.015 &  0.024 & $-$0.012 & $-$0.033 &  0.000 & $-$0.111 \\
HD210918\tablefootmark{*} &5748 &4.319 &$-$0.095 &  1.07 &  0.054 &  0.146 & $-$0.057 &  0.072 &  0.102 &  0.029 &  0.032 &  0.043 &  0.080 & $-$0.012 & $-$0.029 &  0.034 & $-$0.077 \\
HD220507\tablefootmark{*} &5690 &4.247 & 0.013 &  1.07 &  0.102 &  0.159 &  0.036 &  0.126 &  0.153 &  0.071 &  0.084 &  0.051 &  0.114 &  0.004 &  0.017 &  0.121 & $-$0.085 \\
HD222582 &5784 &4.361 &$-$0.004 &  1.07 & $-$0.001 &  0.043 &  0.014 &  0.037 &  0.058 &  0.028 & $-$0.021 &  0.015 &  0.041 &  0.000 &  0.005 &  0.025 & $-$0.065 \\

\noalign{\smallskip}
\hline
\end{tabular}
\tablefoot{
\tablefoottext{*}{$\alpha$-enhanced star}.
}

\end{table*}
}% end of onltab

The estimated 1-sigma errors of the differential abundance ratios 
are given in Table \ref{table:errors}.
Column three lists the error corresponding to the uncertainty of the
atmospheric parameters ($\sigma (\teff ) = \pm 6$\,K, $\sigma (\logg ) = \pm 0.012$,
$\sigma (\feh ) = 0.006$\,dex and $\sigma (\turb ) = 0.017$\,\kmprs ).
For some of the abundance ratios, this error is very
small, because the derived abundances of element X and Fe have 
almost the same dependence on the parameters.
Column four lists the error arising from the measurement 
of equivalent widths, calculated as the line-to-line scatter of the abundances
derived divided by $\sqrt{N_{\rm lines} - 1}$, where $N_{\rm lines}$ 
is the number of spectral lines
applied. In the case of oxygen for which only the weak \oI\ $\lambda 6300.3$ line
was used, $\sigma_{\rm EW}$ was calculated by assuming that the error of the
equivalent width measurement is $\sigma (EW) = \pm 0.2$\,m\AA . Finally, I have
taken into account that differential 3D effects may introduce errors in
the abundance ratios up to 0.005\,dex. Adding these error sources in
quadrature leads to the total adopted errors listed in the last column of 
Table \ref{table:errors}.  

\begin{table}
\begin{center}
\caption[ ]{Uncertainties of abundance ratios.}
\label{table:errors}
\setlength{\tabcolsep}{0.30cm}
\begin{tabular}{rrrrr}
\noalign{\smallskip}
\hline\hline
\noalign{\smallskip}
       & $N_{\rm lines}$ & $\sigma_{\rm atm.par.}$ & $\sigma_{\rm EW}$ & $\sigma_{\rm adopted}$ \\
       &  &  [dex] &  [dex] &  [dex]  \\
\noalign{\smallskip}
\hline
\noalign{\smallskip}
 \cfe  &  2 & 0.009 & 0.009 &  0.013 \\
 \ofe  &  1 & 0.008 & indv. &  indv. \\
 \nafe &  2 & 0.002 & 0.004 &  0.007 \\
 \mgfe &  2 & 0.002 & 0.010 &  0.011 \\
 \alfe &  2 & 0.002 & 0.006 &  0.008 \\
 \sife & 10 & 0.004 & 0.002 &  0.007 \\
 \sfe  &  4 & 0.009 & 0.012 &  0.016 \\
 \cafe &  7 & 0.002 & 0.003 &  0.006 \\
 \tife &  3 & 0.007 & 0.007 &  0.010 \\
 \crfe & 10 & 0.001 & 0.003 &  0.006 \\
 \nife & 14 & 0.001 & 0.003 &  0.006 \\
 \znfe &  3 & 0.004 & 0.009 &  0.011 \\
 \yfe  &  3 & 0.007 & 0.005 &  0.010 \\
\noalign{\smallskip}
\hline
\end{tabular}
\end{center}
\end{table}

\section{Stellar ages}
\label{sect:ages}

As discussed in Sect. \ref{sect:discussion},
it is interesting to have information on stellar ages.
Although the stars belong to the main sequence,
relative ages may be derived from 
the precise values of \teff\ and \logg\ as shown 
in Fig. \ref{fig:iso}, where the stars are compared
to a Yonsei-Yale set of isochrones (Yi et al. \cite{yi01};
Kim et al. \cite{kim02}). 
Using the IDL program made available by Kim et al. to calculate 
isochrones for the actual combination of \feh\ and \alphafe\ for a given
star, it is possible to interpolate to the stellar age and 
its internal 1-sigma error. The results are given in Table \ref{table:ages}
together with stellar masses derived from \teff\ and \logg . The internal error 
of the mass is on the order of $\pm 0.01 M_{\rm sun}$.   

\begin{figure}
\resizebox{\hsize}{!}{\includegraphics{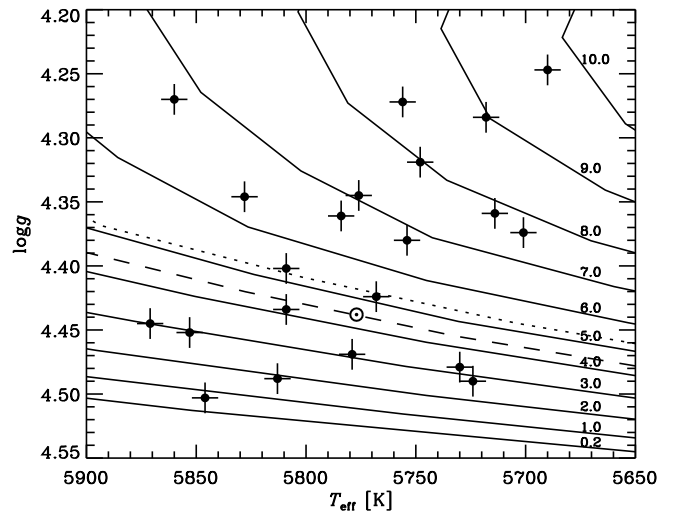}}
\caption{The \teff -- \logg\ diagram of solar twin stars
in comparison with Yonsei-Yale isochrones corresponding to ages
given to the right in Gyr. The isochrones refer to \feh = 0.0 and \alphafe = 0.0,
but 5\,Gyr isochrones corresponding to $\feh = -0.05$
(dashed line) and $\alphafe = +0.1$ (dotted line) are also shown.
The position of the Sun with an age of 4.55\,Gyr is marked.} 
\label{fig:iso}
\end{figure}

\begin{figure}
\resizebox{\hsize}{!}{\includegraphics{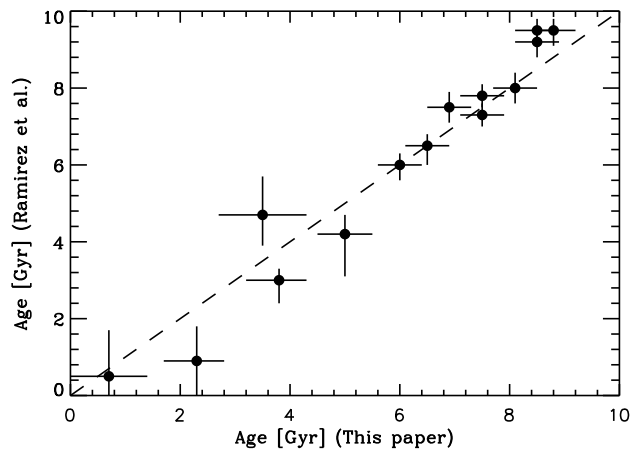}}
\caption{Comparison of stellar ages derived in this paper with
ages determined by Ram\'{\i}rez et al. (\cite{ramirez14b}).}
\label{fig:ages.compare}
\end{figure}

\begin{table}
\begin{center}
\caption[ ]{Stellar ages and masses.}
\label{table:ages}
\setlength{\tabcolsep}{0.30cm}
\begin{tabular}{lccc}
\noalign{\smallskip}
\hline\hline
\noalign{\smallskip}
  Star & Age &  Mass & Note\tablefootmark{a} \\
       & [Gyr] &  [$M_{\odot}] $ & \\
\noalign{\smallskip}
\hline
\noalign{\smallskip}
    HD\,2071 & $3.5\; \pm 0.8$& 0.97& \\
    HD\,8406 & $4.2\; \pm 0.8$& 0.96& \\
   HD\,20782 & $7.5\; \pm 0.4$& 0.97& $1.8 M_{\rm J}$\tablefootmark{b} \\
   HD\,27063 & $2.6\; \pm 0.6$& 1.04& \\   
   HD\,28471 & $7.0\; \pm 0.4$& 0.97& \\
   HD\,38277 & $7.3\; \pm 0.4$& 1.01& \\
   HD\,45184 & $2.7\; \pm 0.5$& 1.06& $0.04 M_{\rm J}$\tablefootmark{c} \\
   HD\,45289\tablefootmark{*} & $8.5\; \pm 0.4$& 1.00& \\
   HD\,71334 & $8.1\; \pm 0.4$& 0.94& \\ 
   HD\,78429 & $7.5\; \pm 0.4$& 1.04& \\
   HD\,88084 & $6.0\; \pm 0.6$& 0.96& \\
   HD\,92719 & $2.7\; \pm 0.6$& 0.99& \\
   HD\,96116 & $0.7\; \pm 0.7$& 1.05& \\
   HD\,96423 & $6.0\; \pm 0.4$& 1.03& \\
  HD\,134664 & $2.3\; \pm 0.5$& 1.07& \\
  HD\,146233 & $3.8\; \pm 0.5$& 1.04& \\
  HD\,183658 & $5.0\; \pm 0.5$& 1.03& \\
  HD\,208704 & $6.9\; \pm 0.4$& 0.98& \\
  HD\,210918\tablefootmark{*} & $8.5\; \pm 0.4$& 0.96& \\
  HD\,220507\tablefootmark{*} & $8.8\; \pm 0.4$& 1.01& \\
  HD\,222582 & $6.5\; \pm 0.4$& 1.01& $7.8 M_{\rm J}$\tablefootmark{d} \\
\noalign{\smallskip}
\hline
\end{tabular}
\tablefoot{
\tablefoottext{*}{$\alpha$-enhanced star}.
\tablefoottext{a}{Mass of detected planet in units of Jupiter's mass}.
\tablefoottext{b}{Jones et al. (\cite{jones06})}.
\tablefoottext{c}{Mayor et al. (\cite{mayor11})}.
\tablefoottext{d}{Butler et al. (\cite{butler06})}.
}

\end{center}
\end{table}

As seen from Fig. \ref{fig:ages.compare}, the stellar ages agree
very well with the values of Ram\'{\i}rez et al. (\cite{ramirez14b}),
who also determined  ages from Yonsei-Yale isochrones in the \teff -- \logg\
diagram. The small systematic deviation for the
three oldest stars may be due to the fact that they have enhanced
\alphafe\ values ($\alphafe \simeq 0.08$),
which were taken into account when deriving
their ages in the present paper, but apparently not by 
Ram\'{\i}rez et al. (\cite{ramirez14b}).

\section{Discussion}
\label{sect:discussion}

\subsection{\xfe\ - stellar age correlations}
\label{sect:abun.var}

\begin{figure*}
\resizebox{\hsize}{!}{\includegraphics{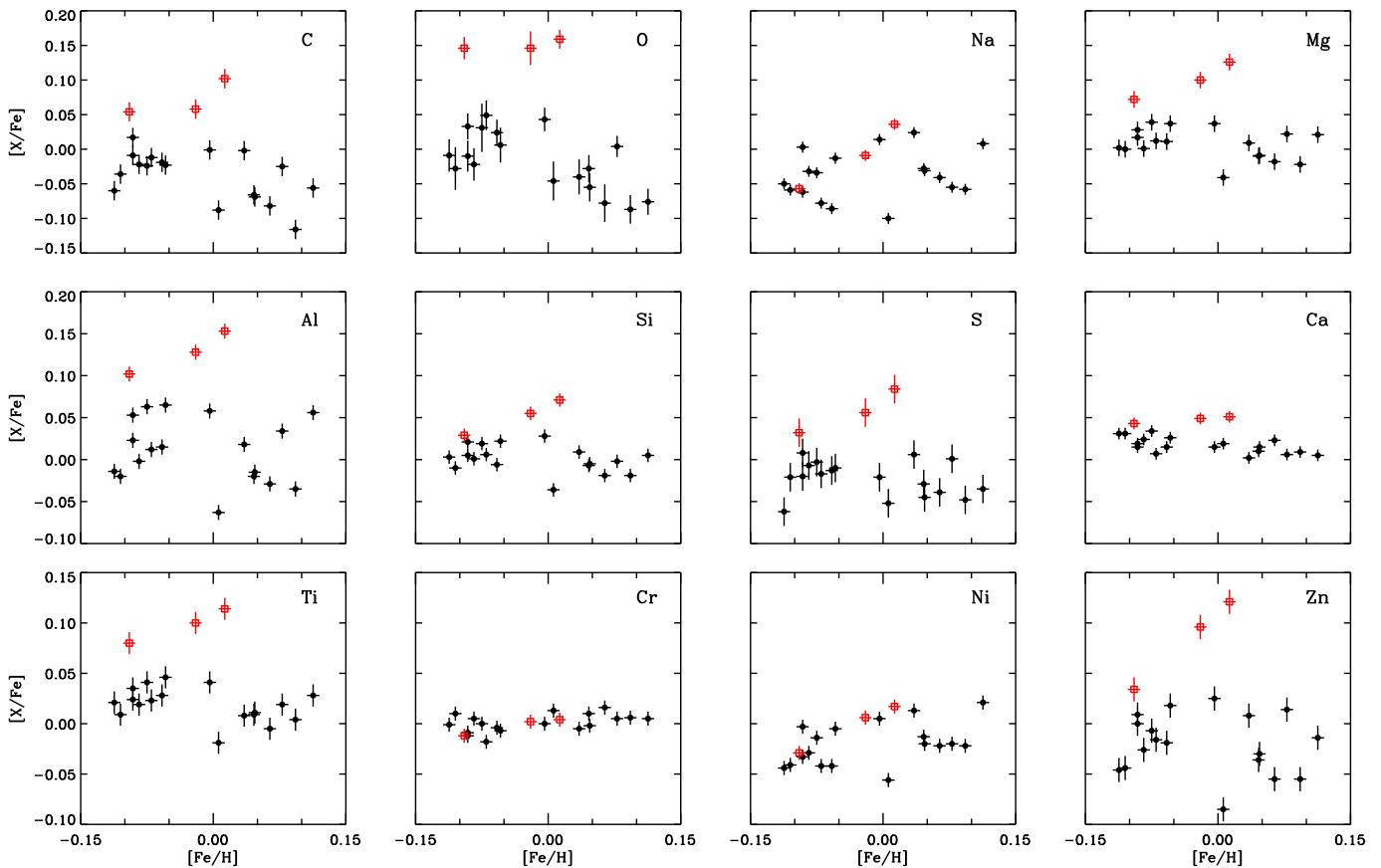}}
\caption{Abundance ratios \xfe\ as a function of \feh . The error bars shown
correspond to the errors listed in Table \ref{table:errors}.
Open (red) squares: $\alpha$-enhanced stars. Filled circles: thin-disk stars}.
\label{fig:xfe-feh}
\end{figure*}

As seen from Fig. \ref{fig:xfe-feh}, the dispersion of \xfe\ at a given \feh\ is larger
than expected from the estimated error bars except in the case of \crfe .
Three stars (\object{HD\,45289}, \object{HD\,210918}, and
\object{HD\,220507}) marked by open (red) squares stand out by
having high \xfe\ ratios of the alpha-capture elements O, Mg, Si, S, and Ti,
and are also high in C, Al, and Zn. These stars probably belong
to the class of metal-rich $\alpha$-enhanced stars first identified
by Adibekyan et al. (\cite{adibekyan11}; \cite{adibekyan12}) as a
population having $\alphafe \simeq +0.1$ in contrast to normal thin-disk stars
with $\alphafe \simeq +0.0$.  Bensby et al. (\cite{bensby14}) confirm the
existence of such metal-rich $\alpha$-enhanced stars, and Haywood et al. (\cite{haywood13})
show that they belong to a thick-disk sequence
in the \alphafe\ - \feh\ diagram for which the ages range from
$\sim \! 13$\,Gyr at $\feh = -1$ to $\sim \! 8$\,Gyr at $\feh = +0.2$. The ages
of the three $\alpha$-enhanced stars found in this paper (8 - 9\,Gyr) agree
with this interpretation. Furthermore, two of the $\alpha$-enhanced stars 
(\object{HD\,45289} and \object{HD\,210918}), have thick-disk kinematics
according to the Galactic velocity components given in Adibekyan et al.
(\cite{adibekyan12}). The third one has typical thin-disk kinematics
like the rest of the sample.

Even if we exclude the three $\alpha$-enhanced stars, there are still variations
in the abundance ratios that cannot be explained by the estimated errors
of the determinations. Interestingly, there is a clear correlation between
\xfe\ and stellar age
for most of the elements as seen from Fig. \ref{fig:xfe-age}. Linear fits
($\xfe = a + b \,\cdot$\,Age) have been obtained by a maximum likelihood 
program that includes errors in both coordinates. Coefficients $a$ and $b$ are
given in Table \ref{table:fits} together with standard deviations and 
reduced chi-squares of the fits. As seen, \mgfe , \sfe , \tife , and \crfe\
are very well correlated with age ($\chi^2_{\rm red} \sim 1$),
but \cfe , \ofe , \alfe , \sife , \crfe , and \znfe\ also show a clear
dependence on age. Exceptions are \nafe\ and \nife , which will be 
discussed below.

\begin{figure*}
\resizebox{\hsize}{!}{\includegraphics{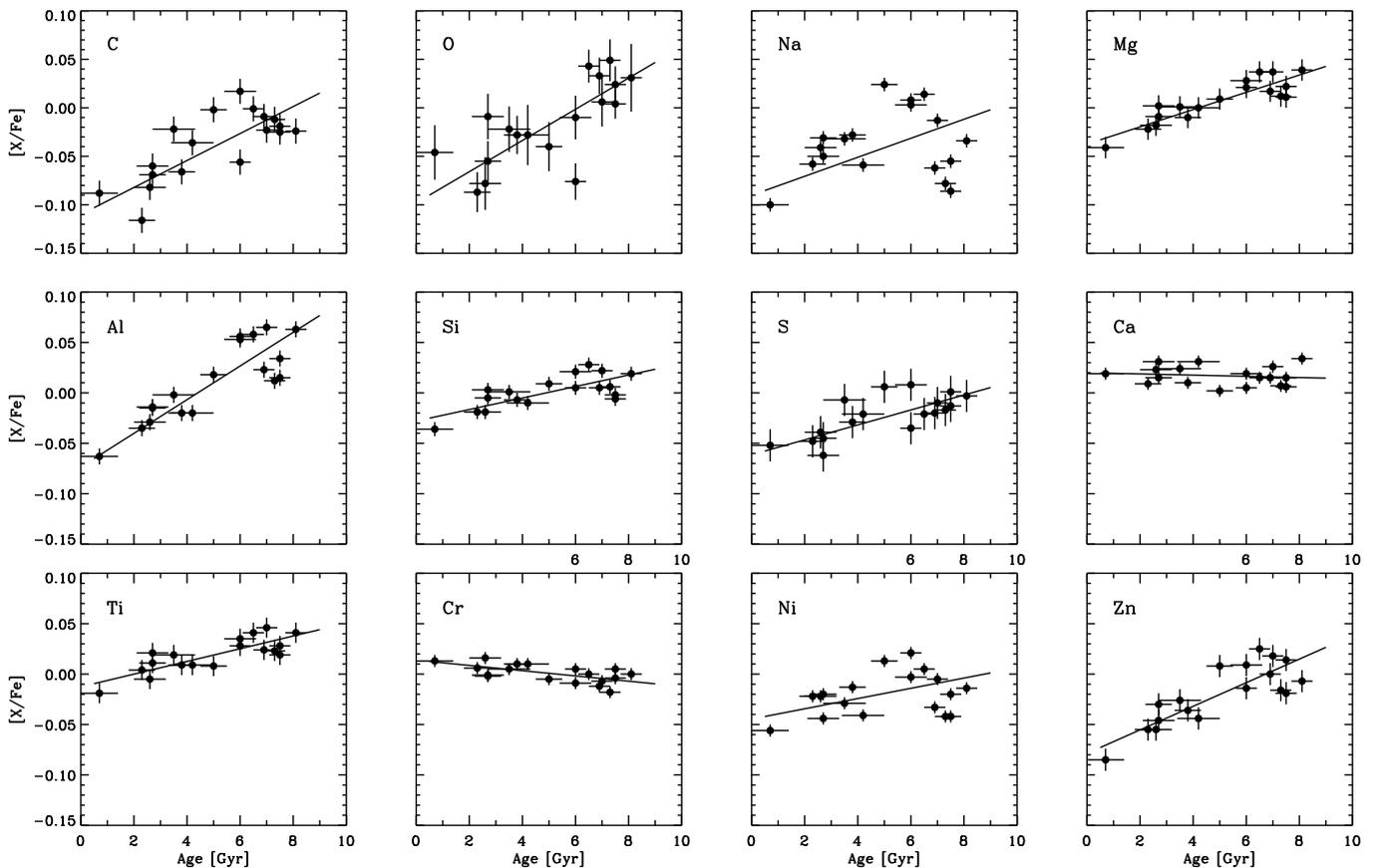}}
\caption{Abundance ratios \xfe\ as a function of stellar age for the thin-disk
stars. The lines show maximum likelihood  linear fits to the data,
with zero points and slope coefficients as given in Table \ref{table:fits}.}
\label{fig:xfe-age}
\end{figure*}

\begin{table}
\begin{center}
\caption[ ]{Linear fits of \xfe\ as a function of stellar age.}
\label{table:fits}
\setlength{\tabcolsep}{0.10cm}
\begin{tabular}{rcccr}
\noalign{\smallskip}
\hline\hline
\noalign{\smallskip}
       & $a$   & $b$                        & $\sigma \xfe$\,\tablefootmark{a}  & $\chi^2_{\rm red}$ \\
       & [dex] & $10^{-3}$\,dex\,Gyr$^{-1}$ &         [dex] &                    \\
\noalign{\smallskip}
\hline
\noalign{\smallskip}
 \cfe  &  $-0.110\,\pm 0.011$ & $+13.9\,\pm 2.0$ & 0.025 & 2.7 \\
 \ofe  &  $-0.098\,\pm 0.015$ & $+16.1\,\pm 2.8$ & 0.030 & 1.9 \\
 \nafe &  $-0.090\,\pm 0.035$ & $ +9.8\,\pm 6.3$ & 0.039 & 20.6 \\
 \mgfe &  $-0.037\,\pm 0.005$ & $ +8.9\,\pm 0.9$ & 0.011 & 0.8  \\
 \alfe &  $-0.074\,\pm 0.011$ & $+16.7\,\pm 1.9$ & 0.020 & 3.6  \\
 \sife &  $-0.028\,\pm 0.005$ & $ +5.7\,\pm 0.9$ & 0.012 & 2.4  \\
 \sfe  &  $-0.058\,\pm 0.007$ & $ +7.0\,\pm 1.3$ & 0.015 & 0.8  \\
 \cafe &  $+0.020\,\pm 0.004$ & $ -0.6\,\pm 0.8$ & 0.010 & 2.7  \\
 \tife &  $-0.013\,\pm 0.005$ & $ +6.3\,\pm 0.8$ & 0.010 & 0.9  \\
 \crfe &  $+0.014\,\pm 0.003$ & $ -2.6\,\pm 0.6$ & 0.007 & 1.4  \\
 \nife &  $-0.044\,\pm 0.011$ & $ +4.8\,\pm 2.0$ & 0.021 &10.7  \\
 \znfe &  $-0.079\,\pm 0.008$ & $+11.7\,\pm 1.4$ & 0.018 & 2.0  \\
 \yfe  &  $+0.146\,\pm 0.011$ & $-33.0\,\pm 2.0$ & 0.023 & 1.6  \\
\noalign{\smallskip}
\hline
\end{tabular}
\tablefoot{
\tablefoottext{a}{Standard deviation of \xfe\ for the linear fit}.}
\end{center}
\end{table}

Figure \ref{fig:feh-mgfe-age} (lower panel) shows how astonishingly well
\mgfe\ for thin-disk stars is correlated with age in contrast to the lack of 
correlation between \feh\ and age (upper panel).
Given that Type Ia Supernovae (SNe) provide about
two-thirds of Fe in the Galactic disk but very little Mg
(Kobayashi et al. \cite{kobayashi06}), the
age correlation of \mgfe\ can be explained by
an increasing number of Type Ia SNe relative to the number of Type II SNe as
time goes on.
The small scatter in the \mgfe - age relation then
indicates that nucleosynthesis products of Type Ia and        
Type II SNe are well mixed locally in the Galactic disk before new
stars are formed. The large scatter in metallicity ($\sigma \feh \simeq 0.07$\,dex)
at a given age may be explained by infall of  
metal-poor gas onto the Galactic disk followed
by star formation before mixing evens out the chemical
inhomogeneities. The infalling gas will mainly contribute hydrogen
causing a decrease of \feh\ but leave \mgfe\ unaffected.
This scenario is discussed in detail by 
Edvardsson et al. (\cite{edvardsson93}), who were first to discover
the large spread in the age-metallicity relation of thin-disk stars in contrast to
a small spread in \alphafe , but their precision of abundance ratios
(typically $\pm 0.05$\,dex)\,\footnote{The Edvardsson et al. sample is not limited to
solar twin stars and abundances were derived from high-resolution spectra with
$S/N \sim 200$.} was not high enough to see a correlation between \alphafe\
and stellar age.
Alternatively, the dispersion in \feh\ at a given
age may be related to orbital mixing of stars born at different
galactocentric distances ($R_{\rm G}$) provided that there 
is a strong radial gradient in \feh\
around the solar distance (Haywood et al. \cite{haywood13}), but then the
very small dispersion in \xfe\ at a given age requires that the star formation
history is independent of $R_{\rm G}$ (Haywood et al. \cite{haywood15}).

\begin{figure}
\resizebox{\hsize}{!}{\includegraphics{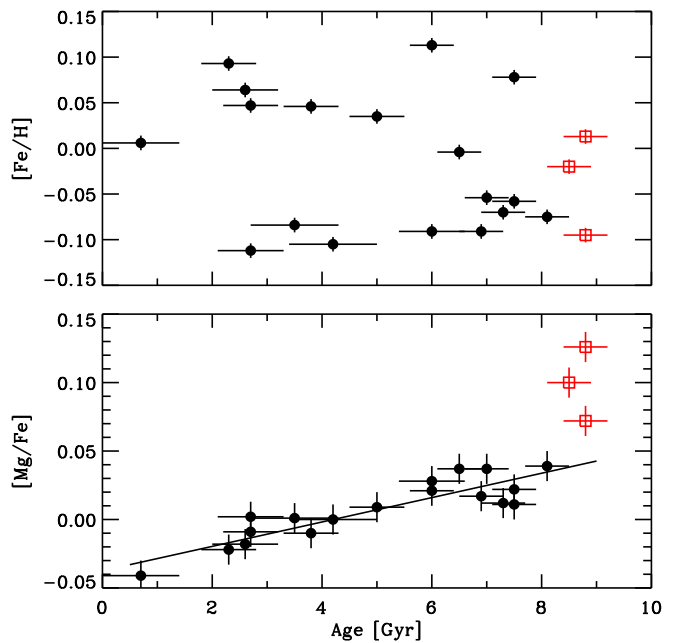}}
\caption{\feh\ and \mgfe\ versus stellar age. Thin-disk stars are shown with
filled circles and $\alpha$-enhanced stars with open (red) squares. 
The line in the lower panel corresponds to the fit given in Table \ref{table:fits}.}
\label{fig:feh-mgfe-age}
\end{figure}

As seen from Fig. \ref{fig:xfe-age} and Table \ref{table:fits}, the slope 
of \xfe\ versus age differs from element to element. To some extent this
may be explained in terms of differences in the relative contributions
of Type Ia and Type II SNe to the various elements, but it cannot explain
that the age-slopes for C, O, and Al are significantly
larger than that of Mg, because for all four elements, the yield of Type Ia SNe 
is negligible compared to the Type II SNe yield (e.g., Kobayashi et al. \cite{kobayashi06}). 
Hence, it seems that one has to include an evolving initial mass function 
to explain the age trends seen in Fig. \ref{fig:xfe-age}.
Asymptotic giant branch (AGB) stars (Karakas \cite{karakas10})
should also be considered, but
cannot solve the problem; they contribute C and Al (as well as Na) 
but no Fe, and hence {\em increase} \xfe\ for these elements as a function of
increasing time, i.e., decreasing stellar age. 

The near-constancy of
\cafe\ as a function of stellar age (see Fig. \ref{fig:xfe-age})
is another puzzling problem. According to yields
of supernovae (Kobayashi et al. \cite{kobayashi06}), one would expect
\cafe\ to increase with age in the same way as \sife\ and \sfe . A similar 
problem is encountered in studies of abundance ratios in early-type
galaxies (Conroy et al. \cite{conroy14}); \cafe\ is constant as
a function of stellar velocity dispersion in
contrast to a rising trend for the other $\alpha$-capture elements.
Mulchaey et al. (\cite{mulchaey14})
suggest that this special behavior of Ca as well as the high Ca abundance
in diffuse gas of clusters of galaxies is due to a new
class of low luminosity supernovae exemplified by \object{SN\,2005E}, for which
the amount of synthesized Ca is 5-10 times greater than that of classical
Type Ia SNe (Perets et al. \cite{perets10}). Perhaps these Ca-rich SNe
are also important for the chemical evolution of the Galactic disk.

Diffusion of elements (gravitational settling and radiative
levitation) should also be considered when discussing
abundances of stars as a
function of time. Helioseismology and models of the Sun indicate that the surface 
abundances of heavy elements (and helium) have decreased by $\sim\!10$\,\% 
relative to their initial values (Christensen Dalsgaard et al.
\cite{jcd96}). According to Turcotte \& Wimmer-Schweingruber
(\cite{turcotte02}), the effect on \xfe\ is, however, much smaller,
i.e., $<\!0.004$\,dex for the elements discussed in this paper.
This means that diffusion can be neglected when discussing
the age trends of \xfe .

The $s$-process element yttrium is not included in Fig.
\ref{fig:xfe-age}, but in contrast to most of the other elements,
\yfe\ decreases with stellar age as previously found for barium 
(Edvardsson et al. \cite{edvardsson93}; Bensby et al. \cite{bensby07}).
This is probably due to an increasing contribution
of Y and Ba from low-mass (1\,-4\, $M_{\rm Sun}$) AGB stars 
as time go on (Travaglio et al. \cite{travaglio04}).
As the slope of \yfe\ versus age is quite steep
($b = -0.033$\,dex Gyr$^{-1}$ with opposite sign to that 
of \mgfe\ ($b = +0.009$\,dex Gyr$^{-1}$), \ymg\ becomes a
sensitive indicator of age as seen from Fig. \ref{fig:ymg-age}.
A maximum likelihood fit to the data yields the relation 
\begin{eqnarray}
\ymg = 0.175 \, (\pm 0.011) - 0.0404 \, (\pm 0.0019) \,\,{\rm Age \, [Gyr]} 
\end{eqnarray}
with $\chi^2_{\rm red} =0.71$.

\begin{figure}
\resizebox{\hsize}{!}{\includegraphics{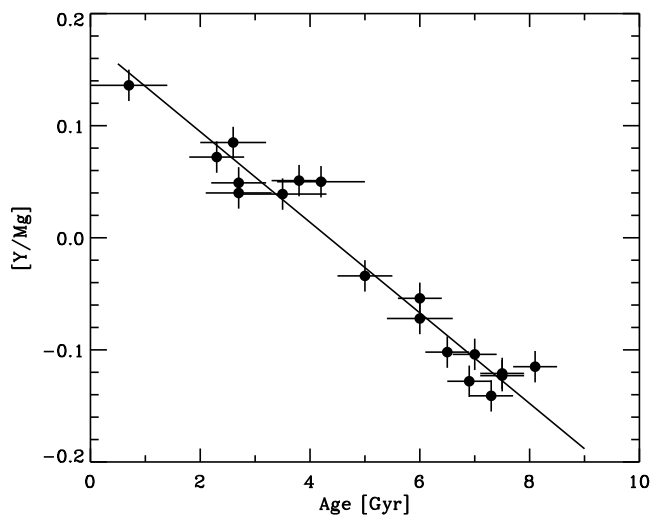}}
\caption{\ymg\ versus stellar age for thin-disk stars.} 
\label{fig:ymg-age}
\end{figure}

For elements having a condensation temperatures markedly different
from that of Fe, the variations in the \xfe -\Tc\ slope discussed
in Sect. \ref{sect:Tc.cor} contribute to the scatter in the
\xfe -age relation. In particular,  this effect may explain why
the scatter in \cfe\ and \ofe\ is larger than expected from the
estimated errors of the determinations, i.e., $\chi^2_{\rm red} = 2.7$
and 1.9, respectively. \co , on the other hand, is not 
much affected by the  \xfe -\Tc\ slope variations, because the two 
elements have similar low condensation temperatures. 

Recently, the C/O ratio\,\footnote{C/O is defined as 
$N_{\rm C}/N_{\rm O}$, where $N_{\rm C}$ and $N_{\rm O}$ are the 
number densities of carbon and oxygen nuclei, respectively, and should not
be confused with the solar-normalized logarithmic ratio, [C/O].}
in solar-type stars has been much discussed,
mainly because its value may influence the composition of planets 
formed in circumstellar disks (e.g., Bond et al. \cite{bond10};
Fortney \cite{fortney12}; Gaidos \cite{gaidos15}). In some studies 
(Delgado Mena et al. \cite{delgado10}; Petigura \& Marcy \cite{petigura11}),  
C/O was found to range from $\sim \! 0.4$ to $\simgt 1.0$,
i.e., up to a factor of two higher than the solar ratio,
C/O$_{\odot} = 0.55 \pm 0.10$ (Asplund et al. \cite{asplund09};
Caffau et al. \cite{caffau11}). 
This has led to speculations about the existence of exoplanets
consisting of carbides and graphite instead of Earth-like silicates
(Bond et al. \cite{bond10}). Alternative determinations of C/O in solar-type stars
(Nissen \cite{nissen13}; Nissen et al. \cite{nissen14}; Teske et al. \cite{teske14}) 
show, however, only a small scatter in C/O and an 
increasing trend as a function of  \feh\ reaching
C/O $\simeq 0.8$ at $\feh \simeq 0.4$. The scatter of C/O
for the solar twin stars is also small, i.e., $\pm 0.035$ only
around a mean value of $<$C/O$>$ = 0.52 (see Fig. \ref{fig:co.ratio-age}).
Furthermore, there is not much evolution of C/O as a function of time
despite the fact that both C/Fe and O/Fe evolve significantly.
This agrees very well with the recent Galactic chemical evolution calculations of
Gaidos (\cite{gaidos15}) as seen from Fig. \ref{fig:co.ratio-age}.
In his model, the dominant contribution of C and O
comes from massive stars, and C/O has only a moderate rise
during the first $\sim5$\,Gyr due to an increasing contribution of carbon 
from stars of lower mass.

\begin{figure}
\resizebox{\hsize}{!}{\includegraphics{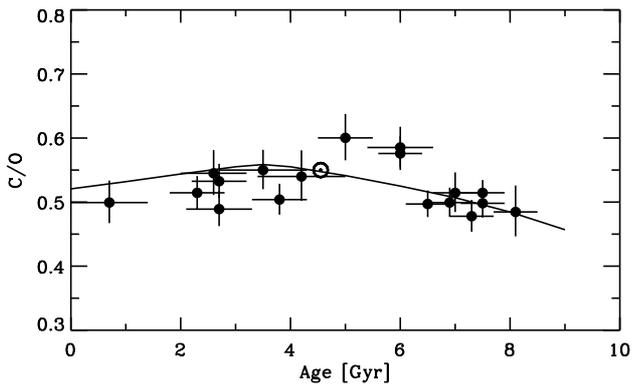}}
\caption{The C/O ratio as a function of stellar age. The line shows the
evolution of C/O (normalized to a solar ratio of 0.55)
according to the chemical evolution calculations of Gaidos (\cite{gaidos15}).}
\label{fig:co.ratio-age}
\end{figure}

As mentioned above, \nafe\ and \nife\ are not well correlated with stellar age;
according to Table \ref{table:fits} the linear fits have $\chi^2_{\rm red} > 10$.
Curiously, there is, however, a tight correlation between \nafe\ and \nife\ as shown
in Fig. \ref{fig:nife-nafe}. A maximum likelihood linear
fit with errors in both coordinates provides the relation
\begin{eqnarray}
\nife = 0.002 \, (\pm 0.002) + 0.580 \, (\pm 0.032) \,\,\nafe
\end{eqnarray}
with a standard deviation $\sigma \nife = 0.008$, which is close to that
expected from the estimated errors of \nafe\ and \nife , i. e., $\pm 0.007$
and $\pm 0.006$\,dex, respectively. 

Apparently, correlated variations of \nife\ and \nafe\ among
solar-metallicity stars have not been noted before. This is understandable
given that the amplitude is small, i.e., $\sim\!0.07$\,dex in \nife ; 
it requires very high
abundance precision to see the variations\,\footnote{For the solar twin stars
studied, abundance ratios derived from HARPS spectra by
Neves et al. (\cite{neves09}) show in fact correlated variations in \nife\ and \nafe\ 
similar to Fig. \ref{fig:nife-nafe}, but it was not noted in that paper nor
in subsequent papers based on HARPS spectra.}.
A Ni-Na relation has, however, been found for thick-disk and halo stars
with $-1.6 < \feh \ -0.4$ (Nissen \& Schuster \cite{nissen97}, \cite{nissen10},
\cite{nissen11}) and for stars in dwarf spheroidal galaxies
in the same metallicity interval (Venn et al. \cite{venn04};
Letarte \cite{letarte10}; Lemasle \cite{lemasle14}). For these metal-poor
stars, the amplitude
of the variations is larger ($\sim\!0.3$\,dex in \nife ), but the slope of the  
\nife\ - \nafe\ relation is about the same as found for the solar-metallicity
stars, i. e. $\Delta \nife / \Delta \nafe \simeq 0.5$.
Furthermore, the lowest Ni/Fe and Na/Fe ratios are found among stars
with the lowest $\alpha$/Fe ratios, which agrees with the trends seen for 
the solar twin stars.

As discussed by Venn et al. (\cite{venn04}), the Ni-Na correlation may be
due to the fact that the yields of Na and the dominant isotope of Ni (Ni$^{58}$)
produced in massive stars exploding as Type II SNe  
both depend on the neutron excess, which itself 
depends on metallicity and the $\alpha$/Fe ratio. Neutrons
are released by the  $^{22}$Ne($\alpha$,n)$^{25}$Mg reaction, where $^{22}$Ne
comes from double $\alpha$-capture on $^{14}$N made from initial C and O
in the CNO cycle. It is, however, difficult to understand why \nafe\ and
\nife\ are not well correlated with stellar age like the other abundance ratios;                 
it was argued above that $\alpha$-capture elements produced
in Type II SNe must be efficiently mixed in interstellar gas
to explain the small scatter of \mgfe\ at a given age, so why
should there be variations of the neutron excess?

\begin{figure}
\resizebox{\hsize}{!}{\includegraphics{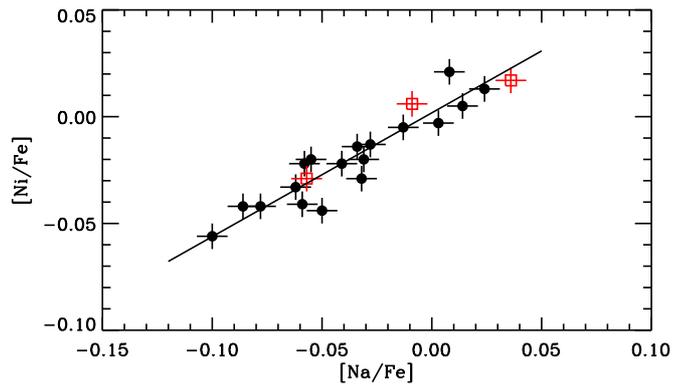}}
\caption{\nife\ versus \nafe\ with the same symbols as used in
Fig. \ref{fig:xfe-feh}. The line corresponds to the
linear fit of \nife\ as a function of \nafe\  given in Eq. (2).}
\label{fig:nife-nafe}
\end{figure}

\subsection{\xfe -\Tc\ correlations}
\label{sect:Tc.cor}

As discussed in Sect. \ref{sect:introduction}, sequestration of refractory 
elements in terrestrial planets or on interstellar dust particles may
introduce star-to-star differences of abundance ratios, which are
correlated with elemental condensation temperature \Tc\ (Lodders \cite{lodders03}).
In order to see if this is the case, \xfe\ has been plotted as a function of
\Tc\ for all stars and linear-least squares fits with 
\xfe\ weighted by the inverse square of its error have been obtained.
The coefficients  
and the reduced $\chi ^2$ of the fits are given in Table \ref{table:slopes}.
Figure \ref{fig:xfe-Tc} shows some typical examples\footnote{Yttrium is not
included in the \xfe -\Tc\ fits, because for many stars \yfe\ has a large deviation.}.

\begin{table}
\caption[ ]{Linear fits, $\xfe = a + b \cdot \Tc$.}  
\label{table:slopes}
\setlength{\tabcolsep}{0.30cm}
\begin{tabular}{lccrc}
\noalign{\smallskip}
\hline\hline
\noalign{\smallskip}
  Star & $a$ & $b$ & $\chi^2_{\rm red}$  \\
       & [dex] &  [$10^{-5}$dex\,K$^{-1}$] &       \\
\noalign{\smallskip}
\hline
\noalign{\smallskip}
    HD\,2071 & $-0.043\; \pm 0.019$ &  $3.1\; \pm 1.5$  &  4.9     \\
    HD\,8406 & $-0.068\; \pm 0.030$ & $4.4\; \pm 2.3$ & 11.2     \\
   HD\,20782 & $-0.060\; \pm 0.034$ & $3.7\; \pm 2.6$ & 16.3      \\
   HD\,27063 & $-0.084\; \pm 0.023$ & $5.5\; \pm 1.8$ &  6.8     \\
   HD\,28471 & $-0.041\; \pm 0.023$ & $4.3\; \pm 1.8$ &  6.8     \\
   HD\,38277 & $-0.051\, \pm 0.034$ & $2.9\; \pm 2.6$ & 14.8     \\
   HD\,45184 & $-0.069\; \pm 0.013$ & $4.6\; \pm 1.0$ &  2.2   \\
   HD\,45289\tablefootmark{*} & $+0.021\; \pm 0.052$ & $1.9\; \pm 4.0$ & 35.1     \\
   HD\,71334 & $-0.059\; \pm 0.026$ &  $5.5\; \pm 2.0$ &  8.3     \\
   HD\,78429 & $-0.038\; \pm 0.024$ & $2.8\; \pm 1.8$ &  8.4     \\
   HD\,88084 & $-0.011\; \pm 0.019$ & $1.9\; \pm 1.5$ &  5.0     \\
   HD\,92719 & $-0.081\; \pm 0.027$ &  $5.4\; \pm 2.1$ &  9.9    \\
   HD\,96116 & $-0.113\; \pm 0.042$ & $6.1\; \pm 3.2$ & 21.7     \\
   HD\,96423 & $-0.060\; \pm 0.015$ & $5.5\; \pm 1.1$ &  3.1     \\
  HD\,134664 & $-0.106\; \pm 0.020$ & $6.7\; \pm 1.6$ &  5.6     \\
  HD\,146233 & $-0.058\; \pm 0.014$ & $3.8\; \pm 1.1$ &  2.7     \\
  HD\,183658 & $+0.003\; \pm 0.012$ & $0.4\; \pm 0.9$ &  1.7     \\
  HD\,208704 & $-0.041\; \pm 0.029$ & $2.7\; \pm 2.2$ & 11.5     \\
  HD\,210918\tablefootmark{*} & $+0.003\; \pm 0.056$ & $1.4\; \pm 4.3$ & 46.2     \\
  HD\,220507\tablefootmark{*} & $+0.085\; \pm 0.053$ &$-1.7   \pm 4.1$ & 43.9     \\
  HD\,222582 & $-0.002\; \pm 0.019$ & $1.7\; \pm 1.5$ &  5.4    \\
\noalign{\smallskip}
\hline
\end{tabular}
\tablefoot{
\tablefoottext{*}{$\alpha$-enhanced star}.
}

\end{table}

As seen from Table \ref{table:slopes}, the reduced $\chi^2$ of the 
fits is larger than one for all stars
indicating that there is never a perfect correlation between \xfe\ and \Tc .
Nevertheless, for several stars the slope
coefficient is more than three times larger than the error of the slope.
The first star, \object{HD\,134664}, in Fig. \ref{fig:xfe-Tc} with
a slope  $\Delta \xfe /\Delta \Tc = 6.7\pm1.6\,\,\times 10^{-5}$dex\,K$^{-1}$
represents such a case.
It has a ratio between refractory elements with $\Tc > 1200$\,K and the 
volatile elements C and O about 0.08\,dex higher than in the
Sun. Assuming that \object{HD\,134664} has a pristine composition,
this means that the Sun has been depleted in refractory elements
by as much as 20\% as first advocated by Mel\'{e}ndez et al. (\cite{melendez09})
from abundance ratios in 11 solar twins.

\begin{figure}
\resizebox{\hsize}{!}{\includegraphics{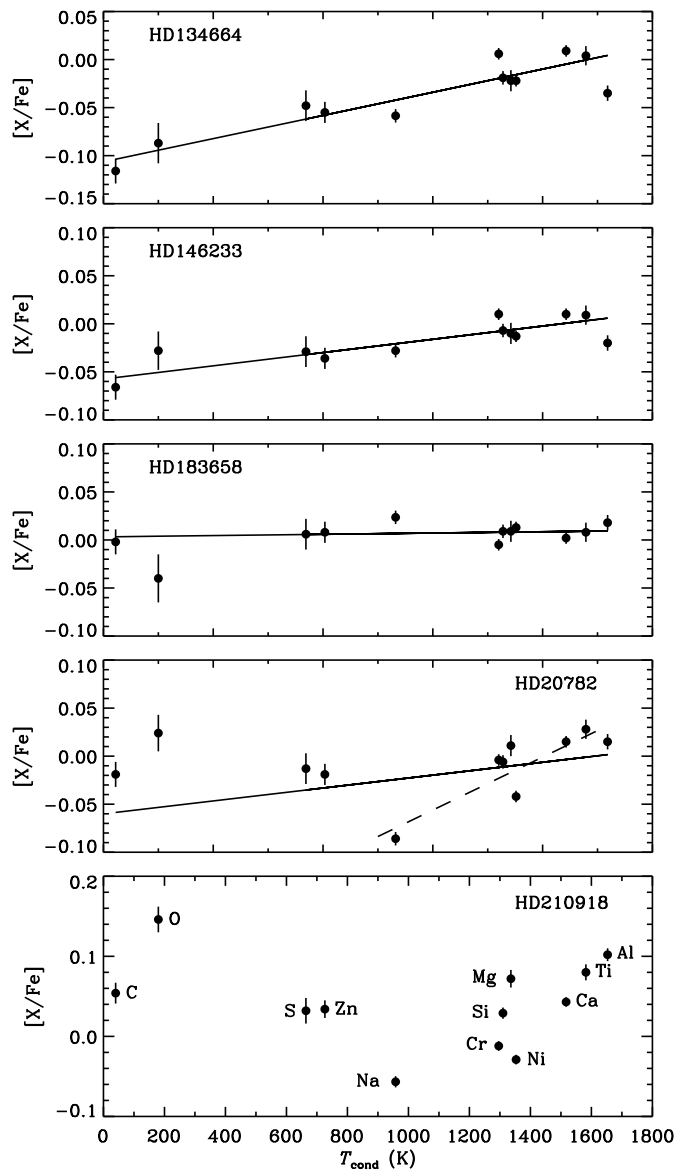}}
\caption{The relation between \xfe\ and elemental condensation temperature
(Lodders \cite{lodders03}) for five representative stars. The lines
show linear fits to the weighted values of \xfe . In the case of
\object{HD\,20782}, the dashed line shows the fit for elements with
$\Tc > 900$\,K.}
\label{fig:xfe-Tc}
\end{figure}

The second star in  Fig. \ref{fig:xfe-Tc}, \object{HD\,146233} (18\,Sco)
represents an intermediate case 
with $\Delta \xfe /\Delta \Tc = 3.8 \pm 1.1\,\,\times 10^{-5}$dex\,K$^{-1}$.
This star has recently been studied in detail by Mel\'{e}ndez et al. (\cite{melendez14a}),
who for 20 elements with $Z \le 30$ find a \Tc -slope of 
$3.5\,\,\times 10^{-5}$dex\,K$^{-1}$ in excellent agreement with the value 
found is this paper\,\footnote{For 13 elements 
in common, the mean deviation and rms scatter for \object{HD\,146233} are
$\Delta \xh $\,(Melendez -- this paper) = $0.003 \pm 0.007$\,dex confirming the
high precision estimated in the two papers.}. 
Neutron capture elements were, however, found to have higher abundances
than expected from the \xfe -\Tc\ relation.
This is confirmed by deriving abundances of Y, Ba and Eu from
the HARPS spectrum of \object{HD\,146233}.

The  third star in Fig. \ref{fig:xfe-Tc}, \object{HD\,183658}, is the
only one in the sample with a well-defined relation between \xfe\ and \Tc\ and a slope
close  to zero. Thus, it is the only star for which the \xfe\
distribution matches that of the Sun, confirming that the Sun is
unusual among solar twin stars.

For the three  $\alpha$-enhanced stars, there is no 
correlation  between \xfe\ and \Tc\ as shown in the case of \object{HD\,210918}
in the last panel of  Fig. \ref{fig:xfe-Tc}. In addition, there are five stars
with $\chi^2_{\rm red} > 10$ for which the correlation is also poor, mainly because
\nafe\ and \nife\ deviate strongly from the mean trend. As shown in
Fig. \ref{fig:xfe-Tc} for the case of \object{HD\,20782} very different slopes
are derived depending on whether one includes all elements 
or fit only elements with a condensation temperature $\Tc > 900$\,K as in 
Ram\'{\i}rez et al. (\cite{ramirez09}). Another solar twin
(\object{HIP\,102152}) with deviating \nafe\ and \nife\ values
was analyzed by Monroe et al. (\cite{monroe13}), who suggest that
the low sodium (and nitrogen) abundance could be due to it's
high age (8\,Gyr). However, stars with low Na and Ni abundances
in this paper cover a broad range in age. Instead, it should
be noted that \object{HIP\,102152} with $\nafe = -0.044$ and
$\nife = -0.026$ according to Monroe et al. fits the Ni-Na relation
shown in Fig. \ref{fig:nife-nafe} almost exactly. 

Recently, Adibekyan et al. (\cite{adibekyan14}) have found evidence for
a dependence of the \xfe -\Tc\ slope on stellar age using data based on HARPS
spectra for 148 solar-like stars with $5600 < \teff < 6375$\,K,
$4.10 < \logg < 4.65$ and $-0.3 < \feh < +0.4$. For this sample,
the \Tc -slope changes by about $6\,\,\times 10^{-5}$dex\,K$^{-1}$ for an
age change of 10\,Gyr. A similar age dependence  of the \xfe -\Tc\ slope
were found for a group of 59 metal-rich solar-analog stars by
Ram\'{\i}rez et al. (\cite{ramirez14a}). As seen from Fig. \ref{fig:slope-age}, there
is also evidence for an age dependence of the slopes derived in this paper.
From a maximum likelihood fit to the data for the thin-disk stars I get  
\begin{eqnarray}
b = 5.6 \, (\pm 0.8) - 0.43 \, (\pm 0.15) \,\, {\rm Age \, [Gyr]} ,
\end{eqnarray}
where $b$ is the \xfe\ versus \Tc\ slope in units of $10^{-5}$dex\,K$^{-1}$. 
The fit has a satisfactory $\chi^2_{\rm red} = 1.5$, but
one star, \object{HD\,183658}, and the Sun have a 3-sigma deviation.
Within the errors, the change of the \Tc -slope with age agrees with the result
of Adibekyan et al.\,\footnote{Note that Adibekyan et al. (\cite{adibekyan14})
define the \Tc -slope with the opposite sign of that used in the present paper.}.
Furthermore, it is noted that the change in \Tc -slope 
is mainly caused by the fact that the change of \cfe\ and \ofe\ with age is steeper
than the corresponding change for refractory elements.

\begin{figure}
\resizebox{\hsize}{!}{\includegraphics{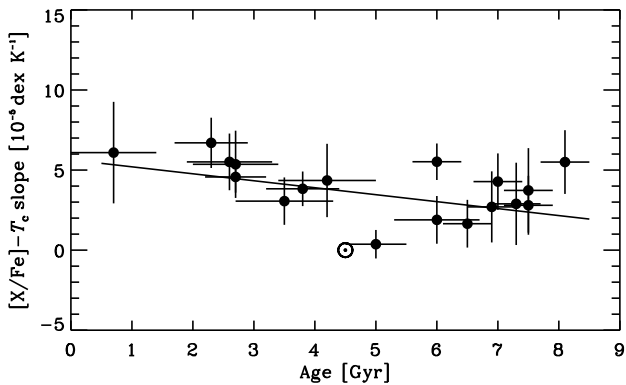}}
\caption{The relation between the \xfe -\Tc\ slope and stellar age 
for thin-disk stars. The line shows the fit given in Eq. (3).} 
\label{fig:slope-age}
\end{figure}

\section{Summary and conclusions}
\label{sect:summary}

In this paper, HARPS spectra with $S/N \ge 600$ have been used to 
derive very precise atmospheric parameters ($\sigma (\teff) \simeq 6$\,K;
$\sigma (\logg) \simeq 0.012$) and elemental abundances
for a sample of 21 solar twin stars in the solar neighborhood.  
Small differential non-LTE effects were taken into account when determining 
\nafe\ ratios and estimated to be negligible for the other abundance
ratios derived. Differential 3D effects seem to be unimportant, although the
impact of possible variations in  the magnetic field strength deserves
further studies. Altogether, the error of the abundance ratios derived
is estimated to be on the order of $\pm 0.01$\,dex, This high precision
is supported by comparing abundances of \object{HD\,146233}    
(18 Sco) with those determined by Mel\'{e}ndez et al. 
(\cite{melendez14a}) from UVES spectra.

In addition, precise stellar ages are obtained by interpolating
between Yonsei-Yale isochrones in the \logg -\teff\ diagram.
A comparison with ages derived by Ram\'{\i}rez et al. (\cite{ramirez14b})
indicates that the ages have internal errors ranging from 0.4 to 0.8\,Gyr.

It is confirmed that the Sun has a lower ratio between refractory and
volatile elements than the large majority of solar twin stars as
found by Mel\'{e}ndez et al. (\cite{melendez09}). Only
one star (\object{HD\,183658}) has the same relation between
\xfe\ and elemental condensation temperature \Tc\ as the
Sun. For several of the other stars, there is a positive
correlation between \xfe\ and \Tc\ with slope coefficients
ranging by more than a factor of two.
This may be explained by various degrees of
sequestration of refractory elements on terrestrial planets
(Mel\'{e}ndez et al. \cite{melendez09}) or by dust-gas segregation
in circumstellar disks (Gaidos \cite{gaidos15}).

For most of the solar twins, the correlation between \xfe\ and \Tc\ 
is rather poor, i.e., $\chi^2_{\rm red} > 1$ for the linear fits.
For many of the elements there is, on the other hand,
an astonishingly tight correlation 
between \xfe\ and stellar age with amplitudes up to 
$\sim \! 0.15$\,dex over an age interval of 8\,Gyr. 
These age correlations may be related to
an increasing number of Type Ia SNe relative to the number of Type II SNe as
time goes on and/or an evolving initial mass function in the Galactic disk. If so,
the slopes of of the age relations for the various elements provide
new constraints on supernovae yields and Galactic chemical evolution.
Furthermore, the small scatter of \xfe\ in the age relations suggests
that nucleosynthesis products of SNe are well mixed locally before
new generations of stars are formed. In contrast, there is a
large scatter of \feh\ at a given age, which may be explained by
infall of metal-poor gas clouds followed by star formation before 
mixing evens out the inhomogeneities (Edvardsson et al. \cite{edvardsson93}).

\cafe\ is nearly constant as a function of stellar age in contrast
to an increasing \xfe\ for the other $\alpha$-capture elements. This
cannot be explained by yields calculated for supernovae of Types II and Ia,
and perhaps it indicates that the new class of low-luminosity Ca-rich 
supernovae discussed by Perets et al. (\cite{perets10}) play an 
important role for the chemical evolution of the Galactic disk.

\ymg\ appears to be a sensitive chronometer for Galactic evolution
as seen from Fig. \ref{fig:ymg-age}.
If \ymg\ can be measured with a precision of 0.04\,dex, the age of solar metallicity
stars can be estimated from Eq. (2) with a precision of 1\,Gyr, but
further studies are needed to see how this relation depends on \feh .
  
The C/O ratio is found to evolve very little over the life time
of the Galactic disk, i.e., to lie within $\pm 0.04$ from the solar
value of C/O $\simeq 0.55$; 
in contrast \cfe\  and \ofe\ change by $\sim \! 0.15$\,dex.
This is of particular interest for discussions of
compositions of exoplanets, because C/O is a key determinant for 
the composition of solids that condense from gas in circumstellar disks.

Na and Ni are exceptions to the tight correlation between \xfe\ and
stellar age, but
interestingly there is a strong correlation between \nife\ and \nafe\ with
a slope of $\simeq 0.6$ and an amplitude of $\sim0.07$\,dex in \nife.
A similar Ni-Na relation has 
previously been found for more metal-poor halo stars (Nissen \& Schuster
\cite{nissen97}, \cite{nissen10}) and
has been ascribed to the fact that the yields of Na and Ni
produced in massive stars and expelled by Type II SNe both
depend on the neutron excess. 
It is, however, difficult to understand why \nafe\ and \nife\ are
not well correlated with stellar age like the other abundance ratios.

The changes of \xfe\ with stellar age and the correlated 
variations of \nife\ and \nafe\ among
solar twin stars complicate the use of the \xfe -\Tc\
slope as a possible signature of terrestrial planets around
stars. Altogether, the interpretation
of abundance ratios in solar twin stars stars, when determined with a precision of
$\sim0.01$\,dex, is difficult, because \xfe\ seems to be 
affected both by  planet formation or dust-gas segregation 
{\em and} chemical evolution in the Galactic disk. Further studies of 
high-precision differential abundances for solar twins and stars
with other metallicities including wide binaries are
needed.

\begin{acknowledgements}
Funding for the Stellar Astrophysics Centre is provided by the
Danish National Research Foundation (Grant agreement no.: DNRF106).
The research is supported by the ASTERISK project
(ASTERoseismic Investigations with SONG and Kepler)
funded by the European Research Council (Grant agreement no.: 267864).
An anonymous referee as well as Vardan Adibekyan, Martin Asplund, 
Karsten Brogaard, Jonay Gonz\'{a}lez Hern\'{a}ndez, Bengt Gustafsson,
Jorge Mel\'{e}ndez, Ivan Ram\'{\i}rez, and David Yong are
thanked for important comments on a first version of the manuscript.
Karin Lind is thanked for IDL programs to calculate non-LTE
corrections for abundances derived from \NaI\ and \FeI\ lines.
This research made use of the SIMBAD database operated
at CDS, Strasbourg, France.
\end{acknowledgements}

\Online

\end{document}